%% file: main.tex
\pdfoutput=1
\documentclass[11pt]{article}
\usepackage{amsmath,amssymb,amsthm,booktabs,array,graphicx,geometry}
\usepackage[numbers,sort&compress]{natbib}
\usepackage{hyperref}
\hypersetup{pdftitle={Elliptic master integrals and the one-loop cosmological wavefunction},pdfauthor={Amara McCune, Matthew D. Schwartz}}
\input{macros}
\newtheorem{proposition}{Proposition}
\begin{document}
\title{Elliptic master integrals \\
and the one-loop cosmological wavefunction}
\author{Amara McCune\\Anthropic STEM Fellow \and Matthew D. Schwartz\\Department of Physics, Harvard University}
\date{}
\maketitle
\begin{abstract}
Loop corrections to the cosmological wavefunction involve similar
integrals to those required for scattering amplitudes. For the one-loop triangle of a conformally coupled scalar, the master integrals are elliptic, and so the triangle wavefunction coefficient was expected to be elliptic as well. Using
a bootstrap approach, we compute this triangle coefficient exactly and show that it is not: the elliptic periods cancel, and the flat-space wavefunction coefficient is a combination of dilogarithms. The cancellation occurs because the residue of the loop integrand on the elliptic sector is a flat-space amplitude, rational in the loop momentum. The one-loop box, by contrast, is not polylogarithmic, even with a single external leg at each site. Its threshold discontinuity involves elliptic integrals, which we also give in closed form. For regular-tetrahedron momenta the periods of a K3 surface enter its loop integrand.
\end{abstract}
\input{sec1_introduction}
\input{sec2_setup}
\input{sec3_elliptic}
\input{sec4_assembled}
\input{sec5_correlators}
\input{sec6_box}
\input{sec7_discussion}

\section*{Acknowledgments}
Many of the computations in this paper were performed by AI: mostly Claude Fable 5.1 (Anthropic) with checks and a few additions by GPT Astra 6 (OpenAI). Much of the text was drafted by AI, then edited and rewritten by the authors. The authors take full responsibility for the validity and accuracy of the results and presentation.

\appendix
\input{appA_method}
\input{appB_box}
\input{appC_derivation}
\bibliographystyle{JHEP}
\bibliography{refs}
\end{document}

%% file: macros.tex
\newcommand{\dd}{\mathrm{d}}
\newcommand{\qG}{q_{\mathcal G}}
\newcommand{\psiflat}{\psi_3}
\newcommand{\Bk}{B}
\newcommand{\Vc}{V}
\newcommand{\Ltwo}{\mathcal L_2}
\newcommand{\Lzero}{\mathcal L_0}
\newcommand{\eps}{\varepsilon}

%% file: sec1_introduction.tex
\section{Introduction}
\label{sec:introduction}

The correlation functions of primordial density perturbations offer one of the most direct windows onto the physics of inflation \cite{Maldacena:2002vr}. Surveys of the cosmic microwave background and of the large-scale distribution of galaxies measure these correlations with ever-increasing precision \cite{Planck:2019kim,DESI:2024mwx,Cabass:2022wjy}, and future observations of the 21-cm line of neutral hydrogen could access far more modes \cite{Loeb:2003ya,Munoz:2015eqa}. They may also carry the imprint of heavy particles present during inflation, as in the cosmological collider program \cite{Chen:2009zp,Arkani-Hamed:2015bza,Lee:2016vti}. The theoretical challenge is to predict them. At tree level and in the simplest models this is now well understood, but beyond the lowest orders of perturbation theory \cite{Weinberg:2005vy,Senatore:2009cf} the relevant integrals have largely resisted computation. At loop level, even the class of functions in which the answer lives remains mostly unknown.

This last question is not merely a technical one. In the study of scattering amplitudes, knowing the function space has proven to be enormously powerful. Once the relevant class of functions is known, an amplitude can often be constructed directly from its singularities and symmetries, without evaluating any integral at all \cite{Dixon:2011pw}. These classes have been mapped out in considerable detail: Feynman integrals evaluate to polylogarithms in the simplest cases \cite{Goncharov:2010jf}, and to integrals over elliptic curves or higher-dimensional varieties in others \cite{Bourjaily:2022bwx}. The cosmological bootstrap \cite{ArkaniHamed:2018kmz,Baumann:2022jpr} aims to construct cosmological correlators in the same way, and it requires precisely the same input. Here the natural analog of the scattering amplitude is the wavefunction of the universe \cite{Hartle:1983ai,Maldacena:2002vr}. Its coefficients (one for each graph of perturbation theory) encode the correlations of the fields at the end of inflation, and equal-time correlators follow from them. The residue of a coefficient at its total-energy pole is the flat-space scattering amplitude \cite{ArkaniHamed:2017fdk,ArkaniHamed:2018kmz}, and in de~Sitter space the wavefunction plays the role of the partition function of a possible dual theory on the late-time boundary \cite{Maldacena:2002vr}. Over the past decade, the computation of these coefficients has increasingly drawn on tools from the amplitudes program, including canonical differential equations \cite{Henn:2013pwa}, integration-by-parts reduction \cite{Chetyrkin:1981qh,Laporta:2000dsw}, and twisted cohomology \cite{Mastrolia:2018uzb,De:2023xue}.

At tree level, the function space is by now well established for conformally coupled scalars in both de~Sitter space and power-law cosmologies. The flat-space wavefunction coefficient is the canonical form of a cosmological polytope \cite{ArkaniHamed:2017fdk}, a rational function with simple poles wherever a partial energy vanishes. The corresponding coefficients in de~Sitter space are polylogarithms, while those in power-law cosmologies are integrals of the same form with non-integer powers of the energies. All of them satisfy differential equations of a universal combinatorial form, known as kinematic flow \cite{ArkaniHamed:2023bsv,ArkaniHamed:2023kig,De:2023xue,Hillman:2019wgh,He:2024olr}.

Much less is known at loop level. The simplest one-loop graph, with two vertices (or sites), is polylogarithmic \cite{Benincasa:2024ptf,Hang:2024xas,Baumann:2024mvm}. For the triangle, ref.~\cite{Benincasa:2024ptf} found that a subset of the master integrals obeys an irreducible second-order differential equation, whose solutions are elliptic integrals. This raises the possibility that elliptic functions enter cosmological wavefunction coefficients already at one loop, in contrast to flat space, where they first appear at two loops, in the sunrise graph \cite{Laporta:2004rb,Bloch:2013tra,Adams:2013nia}.

There is good reason, however, to be cautious about drawing conclusions about an observable from the integrals used to compute it. Master integrals are a convenient basis, but the physical combination can be considerably simpler than its individual pieces, and several earlier results in cosmology are of precisely this kind. The one-loop four-point correlator of a conformally coupled scalar with a quartic interaction in de~Sitter space is expressed in terms of single-valued polylogarithms, because an elliptic sector present in anti-de~Sitter space cancels \cite{Heckelbacher:2022hbq}. More recently, the triangle correlator itself has been computed. The flat-space triangle correlator is a combination of dilogarithms \cite{Pimentel:2026kqc}, as is the triangle correlator of conformally coupled scalars in de~Sitter space \cite{Henn:2026hmy}. These results do not, however, determine the wavefunction coefficient, since correlators lack some of its singularities (those that come from the term of the propagator that imposes the late-time boundary condition) \cite{AguiSalcedo:2023nds,Chowdhury:2023arc}. This leaves a natural question: is the triangle \emph{wavefunction coefficient} itself elliptic, or do the elliptic pieces cancel?

In this paper, we answer this question by computing the one-loop triangle wavefunction coefficient of a conformally coupled scalar directly. To highlight some of our findings (summarized in table~\ref{tab:summary}):
\begin{itemize}
\item The elliptic periods cancel, and the triangle coefficient is a polylogarithmic function of transcendental weight two. With one external leg at each site, it is the combination of three dilogarithms given in eq.~\eqref{eq:closedform}. With several legs per site (the case required by expanding backgrounds), it is given instead by eq.~\eqref{eq:closedformgeneral}, whose dilogarithms have arguments containing ten square roots, all of which become rational with one leg per site. Its counterparts in de~Sitter space and power-law cosmologies follow as integrals of this result over the site energies.
\item The cancellation has a simple origin. The elliptic curve of the master integrals appears on three poles of the loop integrand. At these poles, the residue of the physical integrand is a flat-space amplitude, a rational function of the loop momentum, which does not produce elliptic periods. Individual master integrals contain odd powers of the edge energies, which are square roots in the loop momentum, and so they do see the curve. In their sum, the physical integrand, the square roots cancel and the periods with them (proposition~\ref{prop:residues}). The maximal cuts of the master integrals therefore suggest a larger function space than the coefficient occupies.
\item The same three poles come from the term of the propagator that imposes the late-time boundary condition of the wavefunction. As noted above, such poles are absent from cosmological correlators \cite{AguiSalcedo:2023nds,Chowdhury:2023arc}. In section~\ref{sec:correlators}, we show that the integrand of the triangle correlator has no residue on any line where the loop measure defines an elliptic curve, answering, at the level of the integrand, a question raised in ref.~\cite{Benincasa:2024ptf}. The same holds for the box and the pentagon in $d=N$ dimensions, though not for the box in $d=3$.
\item The box, by contrast, is genuinely not polylogarithmic, even with one leg per site. Here the residue of the loop integrand at the partial-energy pole of a single site involves two square roots rather than one, and together they define an elliptic curve. We give the discontinuity of the box at threshold in closed form in terms of elliptic integrals, and find that for momenta forming a regular tetrahedron the periods of a K3 surface enter the loop integrand. On one line in kinematic space, we also determine the box and its correlator from an exact differential system (section~\ref{sec:box}).
\end{itemize}
\input{table_summary}

Our approach combines exact information about the function space with high-precision numerics. 
This semi-numerical bootstrap approach~\cite{Barrera:2025anr,Acres:2021sss} is well suited to automation, and many of the relevant tools have recently been packaged together in the BootLoops toolkit~\cite{Schwartz:2026bootloops}. BootLoops builds upon the work of a community of mathematical physicists but is effectively
run by a large language model. We used Claude (Anthropic) with BootLoops for many of the computations in this paper. To bootstrap the triangle coefficient, we derive an exact differential operator for it on a line in kinematic space and show that it factors completely into first-order operators, which fixes the function space. We then determine the remaining rational constants from numerical values of the loop integral (section~\ref{sec:assembled}). For general site energies, the symbol fixes the function space and numerical values again fix the constants (section~\ref{sec:offline}). Away from the line, our closed forms rest on their agreement with numerical values of the integral, and for one leg per site we also derive the closed form directly from the integral (appendix~\ref{app:derivation}). This approach is complementary both to the cosmological bootstrap at loop level \cite{Xianyu:2022jwk,Liu:2024str,Liu:2026onb,Westerdijk:2025bnu,Paranjape:2026clb} and to the non-perturbative de~Sitter bootstrap \cite{Hogervorst:2021uvp}, which constrain correlators through symmetries, singularities, and unitarity. Indeed, one of the lessons of this work is that the function space suggested by master integrals can overestimate the true function space of the observable, which is exactly the input a bootstrap requires.

The rest of the paper is organized as follows. In section~\ref{sec:setup}, we define the loop integrand, the loop measure, and the kinematics, and show how coefficients in expanding backgrounds follow from the flat-space wavefunction coefficient. Section~\ref{sec:elliptic} describes the elliptic sector of the triangle, and section~\ref{sec:assembled} shows that its periods cancel and gives the closed forms for the full coefficient. We turn to correlators in section~\ref{sec:correlators} and to the box in section~\ref{sec:box}, and we conclude in section~\ref{sec:discussion}. Appendix~\ref{app:method} details the construction of the closed forms, appendix~\ref{app:box} collects the derivations for the box, and appendix~\ref{app:derivation} sketches a direct derivation of eq.~\eqref{eq:closedform} from the loop integral.

\paragraph{Notation and conventions.} Following the literature, we refer to the vertices of a graph as sites. We call the one-loop graphs with three and four sites the triangle and the box, and more generally the one-loop graph with $N$ sites a polygon. We refer to the magnitude of the momentum of a leg as its energy. When a single external leg is attached to each site, its energy equals the magnitude of its momentum; we refer to this as the case of one leg per site.

%% file: table_summary.tex
\begin{table}[t]
\centering
\small
\renewcommand{\arraystretch}{1.25}
\begin{tabular}{@{}>{\raggedright\arraybackslash}p{0.12\textwidth}>{\raggedright\arraybackslash}p{0.28\textwidth}>{\raggedright\arraybackslash}p{0.27\textwidth}>{\raggedright\arraybackslash}p{0.25\textwidth}@{}}
\toprule
 & Master integrals and loop integrand & Wavefunction coefficient & Correlator \\
\midrule
Triangle & three elliptic sectors with one leg per site (section~\ref{sec:elliptic}), more for general site energies (section~\ref{sec:offline}) & polylogarithmic, of weight two; eqs.~\eqref{eq:closedform} and~\eqref{eq:closedformgeneral} & no residue of the integrand involves an elliptic curve (section~\ref{sec:correlators}) \\
Box, $d=3$ & elliptic curve at the partial-energy pole of a single site (section~\ref{sec:boxresidue}); K3 surface for the regular tetrahedron (section~\ref{sec:boxk3}) & not polylogarithmic, even with one leg per site (section~\ref{sec:boxmonodromy}) & elliptic periods enter on the line of section~\ref{sec:boxline} \\
Pentagon & in $d=3$, genus-five curve at the partial-energy pole of a single site (section~\ref{sec:boxresidue}) & not computed & in $d=5$, no residue of the integrand on a line involves an elliptic curve (section~\ref{sec:census}) \\
\bottomrule
\end{tabular}
\caption{The function spaces of the one-loop polygons of a conformally coupled scalar, as established in this paper. For the triangle, the elliptic curves of the master integrals do not enter the wavefunction coefficient; for the box, they do.}
\label{tab:summary}
\end{table}

%% file: sec2_setup.tex
\section{Setup}
\label{sec:setup}

In this section, we define the wavefunction coefficient $\Vc$ of the one-loop triangle that is the main object of this paper. We begin by recalling how the wavefunction coefficients of a conformally coupled scalar in an expanding background follow from those of flat space, and then give the loop integrand of the triangle, the loop measure, and the external kinematics.

\subsection{Wavefunction coefficients and correlators}
\label{sec:wavefunction}

The late-time state of the fields is described by the wavefunction $\Psi[\phi]$, the probability amplitude for the field to have the spatial profile $\phi_{\vec k}$ at conformal time $\eta_0\to0$, given that it is in the Bunch--Davies vacuum at early times. Expanding its logarithm in powers of the profile,
\begin{equation}
\ln\Psi[\phi] \;=\; -\sum_{n}\frac{1}{n!}\int\prod_{i=1}^{n}\frac{\dd^3k_i}{(2\pi)^3}\;
\psi_n(\vec k_1,\ldots,\vec k_n)\,\phi_{\vec k_1}\cdots\phi_{\vec k_n}\,(2\pi)^3\delta^3\Bigl(\textstyle\sum_i\vec k_i\Bigr)\,,
\label{eq:logpsi}
\end{equation}
defines the wavefunction coefficients $\psi_n$. Equal-time correlators (often called cosmological correlators in the literature) then follow from the Born rule,
\begin{equation}
\langle\phi_{\vec k_1}\cdots\phi_{\vec k_n}\rangle \;=\; \frac{\int\mathcal D\phi\;|\Psi[\phi]|^2\,\phi_{\vec k_1}\cdots\phi_{\vec k_n}}{\int\mathcal D\phi\;|\Psi[\phi]|^2}\,.
\label{eq:born}
\end{equation}
Although wavefunction coefficients and correlators contain the same information, they are different functions of the momenta. Sections~\ref{sec:setup} to~\ref{sec:assembled} concern a wavefunction coefficient, while in section~\ref{sec:correlators} we show that the integrand of the corresponding correlator has fewer singularities.

A key simplification is that a wavefunction coefficient in an expanding background is an integral of the flat-space wavefunction coefficient over energies. To see this, consider a conformally coupled scalar with polynomial self-interactions in a spatially flat Friedmann--Robertson--Walker (FRW) background. After rescaling the field by the scale factor, its free action is that of a massless field in flat space, and the scale factor enters only through a factor $(-\eta)^{-(1+\sigma)}$ at each interaction vertex. Here the exponent $\sigma$ depends on both the background and the interaction; for a cubic interaction in de~Sitter space, $\sigma=0$. After the Wick rotation $\eta=\mathrm i t$ with $t>0$, an external leg of energy $X$ contributes the bulk-to-boundary propagator $e^{-Xt}$, while the vertex factor is $t^{-(1+\sigma)}$ up to a phase. Since
\begin{equation}
t^{-(1+\sigma)} \;=\; \frac{1}{\Gamma(1+\sigma)}\int_0^\infty\dd\omega\;\omega^{\sigma}\,e^{-\omega t}\,,
\label{eq:mellin}
\end{equation}
the factor $e^{-\omega t}$ multiplies $e^{-Xt}$ and replaces the energy $X$ at that vertex by $X+\omega$ \cite{ArkaniHamed:2017fdk,Hillman:2019wgh}. A coefficient in the expanding background is therefore an integral of the flat-space wavefunction coefficient over shifted energies,
\begin{equation}
\psi^{(\sigma)}_n(X;y) \;\propto\; \int_0^\infty\prod_v\dd\omega_v\,\omega_v^{\sigma}\;\psi_n(X+\omega;y)\,,
\label{eq:siteintegrals}
\end{equation}
where $\psi_n$ without a superscript denotes the flat-space wavefunction coefficient. Here $X_v$ is the total energy of the external legs at site $v$ and $y_e$ is the energy of the internal edge $e$, also referred to in the literature as the external and internal energies. Note that in flat space the vertex factor is absent, which is the case $\sigma=-1$. There the representation~\eqref{eq:mellin}, valid for $\sigma>-1$, degenerates, $\omega^\sigma/\Gamma(1+\sigma)\to\delta(\omega)$, and $\psi^{(\sigma)}_n$ reduces to its integrand $\psi_n$. Since every coefficient in a background of this type follows from $\psi_n$, we study the latter directly in sections~\ref{sec:elliptic} and~\ref{sec:assembled}, returning to the integrals of eq.~\eqref{eq:siteintegrals} for de~Sitter space in section~\ref{sec:desitter}.

\subsection{The loop integrand}
\label{sec:integrand}

The flat-space wavefunction coefficients are integrals over the times $t_v>0$, in which each site contributes a factor $e^{-X_vt_v}$ and each internal edge contributes the bulk-to-bulk propagator, which vanishes at the late-time boundary,
\begin{equation}
G_y(t,t') \;=\; \frac{1}{2y}\Bigl[e^{-y|t-t'|}-e^{-y(t+t')}\Bigr]\,.
\label{eq:propagator}
\end{equation}
We omit the coupling constants and an overall phase, so that every $\psi_n$ below is a positive rational function of the energies. The first term in eq.~\eqref{eq:propagator} is simply the time-ordered propagator of flat space. The second term, by contrast, is fixed by the boundary condition at $t=0$ and factorizes into a function of $t$ times a function of $t'$. This second term has no counterpart in a scattering amplitude, and, as we will see, it is the origin of the elliptic curve of section~\ref{sec:elliptic}.

At tree level, every factor in the denominator is a partial energy, which we now define. From here on, the subscript of $\psi$ counts the sites of the graph. A single site with no edges gives $\psi_1=1/X$, while two sites joined by an edge give
\begin{equation}
\psi_2 \;=\; \int_0^\infty\!\dd t_1\,\dd t_2\;e^{-X_1t_1-X_2t_2}\,G_y(t_1,t_2) \;=\; \frac{1}{(X_1+X_2)(X_1+y)(X_2+y)}\,.
\label{eq:twosite}
\end{equation}
Each factor in the denominator is the sum of the $X_v$ of a connected set of sites and of the $y_e$ of the edges that join the set to the remaining sites; this sum is known as a partial energy. The denominator of every tree graph consists of partial energies \cite{ArkaniHamed:2017fdk}. Because the boundary breaks time translations, energy is not conserved, and the coefficient is instead singular wherever a partial energy vanishes.

\begin{figure}[t]
\centering
\includegraphics[width=0.95\textwidth]{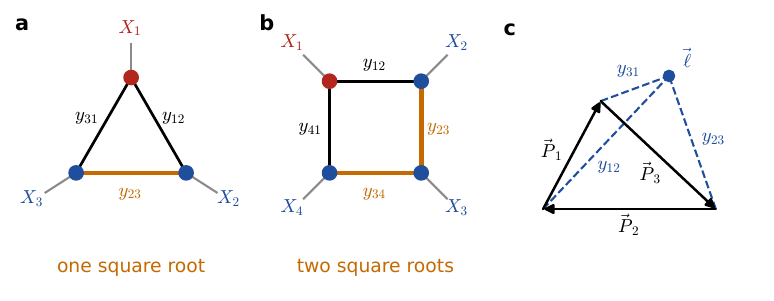}
\caption{(a) The one-loop triangle and (b) the one-loop box. Each site has the energy $X_v$ of its external legs, and each edge has the energy $y_e$. At the partial-energy pole of site~1, marked in red, the residue of the loop integrand is even in the energies of the black edges and contains the square root of a quadratic polynomial for each orange edge (sections~\ref{sec:parity} and~\ref{sec:boxresidue}). (c) The loop measure of the triangle. The external momenta form a triangle with sides $P_v$, and the edge energies are the distances from the loop momentum $\vec\ell$ to its corners. In general $\vec\ell$ does not lie in the plane of the triangle.}
\label{fig:setup}
\end{figure}

We now turn to the one-loop triangle, the graph with three sites and three edges shown in figure~\ref{fig:setup}(a). Its sites have energies $X_1,X_2,X_3$, and the edge between sites $v$ and $w$ has energy $y_{vw}$. Carrying out the time integrals gives
\begin{equation}
\psiflat(X;y) \;=\; \frac{2\,\mathcal N(X;y)}{\qG\,\prod_{i=1}^{9}q_i(X;y)}\,,
\qquad \qG=X_1+X_2+X_3\,,
\label{eq:integrand}
\end{equation}
with ten linear factors in the denominator. In addition to the total energy $\qG$, these are the nine factors $q_1,\ldots,q_9$, which we list row by row,
\begin{equation}
\begin{aligned}
&X_1+y_{12}+y_{31}\,,\quad X_2+y_{12}+y_{23}\,,\quad X_3+y_{23}+y_{31}\,,\\
&X_1+X_2+y_{23}+y_{31}\,,\quad X_2+X_3+y_{12}+y_{31}\,,\quad X_1+X_3+y_{12}+y_{23}\,,\\
&\qG+2y_{12}\,,\quad \qG+2y_{23}\,,\quad \qG+2y_{31}\,.
\end{aligned}
\label{eq:factors}
\end{equation}
The numerator $\mathcal N$ is an irreducible homogeneous quartic polynomial with positive coefficients, which we provide in the ancillary files. We refer to $\psiflat(X;y)$ as the loop integrand. It can also be obtained from a recursion that cuts one edge at a time \cite{ArkaniHamed:2017fdk},
\begin{equation}
\begin{aligned}
\qG\,\psiflat(X;y) \;=\;{}& \psi_{\rm chain}\bigl(X_2+y_{12},X_3,X_1+y_{12};y_{23},y_{31}\bigr)\\
&+\psi_{\rm chain}\bigl(X_3+y_{23},X_1,X_2+y_{23};y_{31},y_{12}\bigr)\\
&+\psi_{\rm chain}\bigl(X_1+y_{31},X_2,X_3+y_{31};y_{12},y_{23}\bigr)\,,
\end{aligned}
\label{eq:recursion3}
\end{equation}
where each term is the coefficient of the chain in which one edge $e$ has been cut, with $y_e$ added to the energies of the two sites at its ends. For reference, the coefficient of the chain of three sites with energies $x_1,x_2,x_3$ and edges $y,y'$ is
\begin{equation}
\psi_{\rm chain}(x_1,x_2,x_3;y,y') \;=\; \frac{(x_1+x_2+y')^{-1}+(x_2+x_3+y)^{-1}}{(x_1+x_2+x_3)(x_1+y)(x_2+y+y')(x_3+y')}\,.
\label{eq:chain}
\end{equation}

The total energy and the first two rows of eq.~\eqref{eq:factors} are partial energies (of the whole graph, of a single site, and of a pair of sites, respectively), but the factors in the last row are not. Instead, the factor $\qG+2y_e$ contains the energy of the edge $e$ twice: it is the total energy of the chain in which $e$ is cut and its two ends become external legs. These factors arise from the second term of the propagator~\eqref{eq:propagator}, and they play a central role in what follows. Every elliptic curve that we find in the triangle involves one of the three factors $\qG+2y_e$, and, as we show in section~\ref{sec:correlators}, these factors are absent from the correlator integrand.

\subsection{The loop measure}
\label{sec:measure}

The edge energies depend on the loop momentum $\vec\ell$ through $y_{12}=|\vec\ell|$, $y_{23}=|\vec\ell+\vec P_2|$ and $y_{31}=|\vec\ell-\vec P_1|$, where $\vec P_v$ is the total momentum entering site $v$ and $\vec P_1+\vec P_2+\vec P_3=0$. Since the integrand depends on $\vec\ell$ only through these three lengths, we use them as integration variables. Geometrically, the three edge energies are the distances from the point $\vec\ell$ to the corners of the triangle formed by the external momenta (figure~\ref{fig:setup}(c)). In three dimensions, three distances fix a point up to reflection through the plane of the triangle, and for each of the two points the Jacobian is $y_{12}\,y_{23}\,y_{31}/(6\,\mathrm{Vol})$, where $\mathrm{Vol}$ is the volume of the tetrahedron with apex $\vec\ell$. With $288\,\mathrm{Vol}^2=\Bk$ and the sum over the two points this gives
\begin{equation}
\dd^3\ell \;=\; 4\sqrt2\;\Bk(y;P)^{-1/2}\;y_{12}\,y_{23}\,y_{31}\;\dd y_{12}\,\dd y_{23}\,\dd y_{31}\,,
\label{eq:measure}
\end{equation}
where $\Bk$ is the Cayley--Menger determinant of that tetrahedron,
\begin{equation}
\Bk(y;P) \;=\; \det\begin{pmatrix}
0&1&1&1&1\\
1&0&P_2^2&P_1^2&y_{12}^2\\
1&P_2^2&0&P_3^2&y_{23}^2\\
1&P_1^2&P_3^2&0&y_{31}^2\\
1&y_{12}^2&y_{23}^2&y_{31}^2&0
\end{pmatrix},
\label{eq:CM}
\end{equation}
and $P_v=|\vec P_v|$. The integration region is $\Bk>0$, where the tetrahedron exists, and the polynomial $\Bk$ is quadratic in each $y_e^2$. Readers familiar with the Feynman-integral literature will recognize this change of variables as the Baikov representation, with $\Bk$ the Baikov polynomial; it is also the representation in which ref.~\cite{Benincasa:2024ptf} studied the triangle.

The main object of this paper is the integral of the loop integrand over the loop momentum, that is, the one-loop wavefunction coefficient of the triangle. We normalize it as
\begin{equation}
\Vc(X;P) \;=\; \int_{\Bk>0}\Bk(y;P)^{-1/2+\eps}\;y_{12}\,y_{23}\,y_{31}\;\qG\,\psiflat(X;y)\;\dd y_{12}\,\dd y_{23}\,\dd y_{31}
\;\;\xrightarrow{\;\eps\to0\;}\;\; \frac{\qG}{4\sqrt2}\int\dd^3\ell\;\psiflat\,.
\label{eq:V}
\end{equation}
The factor $\qG$ removes the pole at vanishing total energy. Since the loop integrand falls off as $|\vec\ell|^{-5}$ at large loop momentum, $\Vc$ is finite in three dimensions. The exponent $\eps$ is the parameter of dimensional regularization, with $d=3+2\eps$; we need it only because the individual master integrals of section~\ref{sec:elliptic} have poles at $\eps=0$, which cancel in $\Vc$. Note that the square root $\Bk^{-1/2}$ is absent at tree level, where the integrands are rational. The elliptic curve of section~\ref{sec:elliptic} requires both this square root and the second term of eq.~\eqref{eq:propagator}.

\subsection{Kinematics}
\label{sec:kinematics}

The coefficient $\Vc$ depends on the site energies $X_v$ and on the momentum magnitudes $P_v$, the latter subject to the triangle inequalities. If a single external leg is attached to each site, its energy equals the magnitude of its momentum, so that
\begin{equation}
X_v \;=\; P_v\,,\qquad v=1,2,3\,.
\label{eq:oneleg}
\end{equation}
With several legs at a site, on the other hand, $X_v$ exceeds $P_v$ unless the legs are collinear.

On a line in kinematic space, the coefficient depends on a single variable and obeys an ordinary differential equation. Following ref.~\cite{Benincasa:2024ptf}, which studies the differential equations along lines of the form $X=(\alpha\lambda,\lambda,1)$, we take $\alpha=2$ and study the line
\begin{equation}
X \;=\; P \;=\; (2\lambda,\,\lambda,\,1)\,,\qquad \tfrac13<\lambda<1\,,
\label{eq:line}
\end{equation}
where the range of $\lambda$ is set by the triangle inequalities. The value of $\alpha$ has no special significance; as a control, we also derived the second-order operator of section~\ref{sec:periods} at $\alpha=3$, where it again agrees with ref.~\cite{Benincasa:2024ptf}. The differential equations of section~\ref{sec:elliptic} and of sections~\ref{sec:L0}--\ref{sec:closedform} are derived on this line; section~\ref{sec:offline} uses a second line with $X_v>P_v$. However, the case of one leg per site does not suffice for expanding backgrounds: the integrals of eq.~\eqref{eq:siteintegrals} shift $X_v$ at fixed $P_v$, so every coefficient in an expanding background requires $\Vc$ at $X_v>P_v$. We treat general site energies in section~\ref{sec:offline}.

%% file: sec3_elliptic.tex
\section{The elliptic sector}
\label{sec:elliptic}

In this section, we describe the elliptic curve of the triangle and the integrals in which it appears. After expressing $\Vc$ through a finite set of master integrals, we derive the curve, its periods, and their differential operator, and then describe the master integrals themselves. We will see that three sectors of master integrals are elliptic. Whether $\Vc$ itself is elliptic is, however, a separate question, which we answer in section~\ref{sec:assembled}.

\subsection{Master integrals}
\label{sec:masters}

The integrand of eq.~\eqref{eq:V} is a polynomial divided by a product of linear factors and multiplied by a power of $\Bk$. The integral of eq.~\eqref{eq:V} therefore belongs to the family
\begin{equation}
I[\mathcal Q;n] \;=\; \int_{\Bk>0}\Bk(y;P)^{-1/2+\eps}\;\frac{\mathcal Q(y)}{\prod_{i=1}^{9}q_i(X;y)^{n_i}}\;\dd^3y\,,
\label{eq:family}
\end{equation}
where $\mathcal Q$ is a polynomial, the $q_i$ are the nine factors of eq.~\eqref{eq:factors}, and the $n_i$ are non-negative integers. In dimensional regularization, the integral of a total derivative over the region $\Bk>0$ vanishes, and the resulting integration-by-parts identities relate the members of the family linearly, with coefficients that are rational functions of the kinematics and of $\eps$. Just as for Feynman integrals, the family is then spanned by a finite number of master integrals, which we group into sectors according to which of the factors $q_i$ appear in the denominator. The derivative of a master integral with respect to $\lambda$ is a combination of master integrals from its own sector and from sectors with fewer denominators. Dropping the latter yields the homogeneous differential equation of the sector, whose solutions determine the function space of that sector.

The sectors built only from the first six factors of eq.~\eqref{eq:factors} produce polylogarithms on the line~\eqref{eq:line}, but the sectors with a factor $\qG+2y_e$ do not. In particular, ref.~\cite{Benincasa:2024ptf} found that the sector with the single denominator $\qG+2y_{12}$ contains master integrals whose homogeneous equation is irreducible and of second order, and the same holds for $\qG+2y_{23}$ and $\qG+2y_{31}$ after a permutation of the sites. We refer to these three sectors as the elliptic sectors. Without loss of generality, sections~\ref{sec:curve} to~\ref{sec:ellipticresults} treat the sector with denominator $\qG+2y_{12}$.

\subsection{The curve}
\label{sec:curve}

The elliptic curve can be read off directly from the loop measure. Recall that an integrand containing the square root of a polynomial $p(x)$ is a differential on the curve $w^2=p(x)$. If $p$ has degree two, the curve has genus zero and the integral is a logarithm; if $p$ has degree three or four with distinct roots, the curve is elliptic, and the integrals of its holomorphic differential over closed paths are its periods.

Since the residue of the integrand at the pole of a sector obeys the homogeneous equation of that sector, in the sector with denominator $\qG+2y_{12}$ we take the residue at $y_{12}=-\qG/2$ (the maximal cut of the sector). This leaves an integral over $y_{23}$ and $y_{31}$ whose integrand depends on these variables only through the square root of $\Bk$. In terms of $z_e=y_e^2$, the polynomial $\Bk$ is quadratic, and along the ray $z_{31}=r\,z_{23}$ its leading behavior at large $z_{23}$ is
\begin{equation}
\Bk \;=\; -A(r)\,z_{23}^2+O(z_{23})\,,\qquad A(r)\;=\;2\bigl[P_2^2\,r^2-\bigl(P_1^2+P_2^2-P_3^2\bigr)\,r+P_1^2\bigr]\,,
\label{eq:At}
\end{equation}
where $A(r)$ does not depend on $y_{12}$. If the rest of the integrand were a rational function of $r$, the square root of the quadratic $A(r)$ would define a curve of genus zero. The crucial point is that the integrands of the elliptic sectors are not functions of the squares $y_e^2$ alone, even though $\Bk$ is: a master integral with unit numerator has the measure $\dd y_{23}\,\dd y_{31}$ rather than $\dd z_{23}\,\dd z_{31}$. Using $\dd y_e=\dd z_e/(2\sqrt{z_e})$, the measure at large $z_{23}$ becomes
\begin{equation}
\Bk^{-1/2}\,\dd y_{23}\,\dd y_{31} \;\longrightarrow\; \frac{\dd z_{23}}{z_{23}}\;\frac{\dd r}{4\sqrt{-r\,A(r)}}\,,\qquad
\frac{\dd r}{\sqrt{r\,A(r)}} \;=\; \frac{2\,\dd\tau}{\sqrt{A(\tau^2)}}\,,\qquad \tau=\frac{y_{31}}{y_{23}}\,.
\label{eq:measurecurve}
\end{equation}
The residue at large $z_{23}$ is thus a differential in $\tau$, and the square root of the quartic $A(\tau^2)$ defines the elliptic curve
\begin{equation}
w^2 \;=\; P_2^2\,\tau^4-\bigl(P_1^2+P_2^2-P_3^2\bigr)\,\tau^2+P_1^2\,.
\label{eq:quartic}
\end{equation}
Since the discriminant of $A(r)$ is proportional to the K\"all\'en function of the $P_v^2$, the four branch points are distinct away from the folded configurations, in which the three momenta are collinear.

Notably, the curve depends only on the momenta $P_v$, and not on the site energies $X_v$. It lies at infinity on the plane $\qG+2y_{12}=0$, where $y_{23}$ and $y_{31}$ are large at fixed $y_{12}$, which is not a region of real loop momentum. An integral therefore depends on the curve only through the behavior of its integrand at large $y_{23}$ and $y_{31}$ on that plane.

On the line~\eqref{eq:line}, the curve is isomorphic to $w^2=x(1-x)(1-m\,x)$, where the parameter $m$ is the cross-ratio of the four branch points,
\begin{equation}
m(\lambda) \;=\; \frac{\lambda^2-1}{9\lambda^2-1}\,,
\label{eq:modulus}
\end{equation}
in agreement with ref.~\cite{Benincasa:2024ptf}. The sectors with denominators $\qG+2y_{23}$ and $\qG+2y_{31}$ have curves of the same form with the momenta permuted, though the three curves are not isomorphic to one another.

\subsection{Periods and their differential equation}
\label{sec:periods}

The homogeneous equation of the elliptic sector is the differential equation obeyed by the periods, whose operator is known as the Picard--Fuchs operator. On the line~\eqref{eq:line}, it is given by \cite{Benincasa:2024ptf}
\begin{equation}
\Ltwo \;=\; \frac{\dd^2}{\dd\lambda^2}
+\frac{45\lambda^4-30\lambda^2+1}{\lambda\,(\lambda^2-1)(9\lambda^2-1)}\,\frac{\dd}{\dd\lambda}
+\frac{27\lambda^2-10}{(\lambda^2-1)(9\lambda^2-1)}\,,
\label{eq:L2}
\end{equation}
whose solutions are complete elliptic integrals of the first kind, $K(m)=\int_0^{\pi/2}(1-m\sin^2\theta)^{-1/2}\,\dd\theta$, multiplied by an algebraic prefactor. On the physical interval $\tfrac13<\lambda<1$, the two periods are
\begin{equation}
\varpi_0(\lambda) \;=\; \frac1\lambda\,K\Bigl(\frac{1-\lambda^2}{8\lambda^2}\Bigr)\,,\qquad
\varpi_1(\lambda) \;=\; \frac1\lambda\,K\Bigl(\frac{9\lambda^2-1}{8\lambda^2}\Bigr)\,,
\label{eq:periods}
\end{equation}
where the argument of the first is $m/(m-1)$, with $m(\lambda)$ the parameter of eq.~\eqref{eq:modulus}. The first period is analytic at $\lambda=1$ and has a logarithmic singularity at $\lambda=\tfrac13$, while the second has the opposite behavior. The operator $\Ltwo$ has regular singular points at $\lambda=0,\pm\tfrac13,\pm1$ and $\infty$, and at each finite singular point the two local exponents coincide, so that one solution is analytic there and the other has a logarithm. Both ends of the physical interval $\tfrac13<\lambda<1$ are singular points of this kind. This fact will be useful in section~\ref{sec:assembled}, since any function that contains the periods~\eqref{eq:periods} inherits their logarithmic singularities.

We note that the operator $\Ltwo$ is the same for the denominators $\qG+2y_{12}$ and $\qG+y_{12}$, since the curve~\eqref{eq:quartic} does not depend on the position of the pole in $y_{12}$. Ref.~\cite{Benincasa:2024ptf} writes $\qG+y_{12}$ in its section on the triangle, whereas the time integrals of section~\ref{sec:integrand} give $\qG+2y_{12}$, which we use throughout.

\subsection{The master integrals of the elliptic sector}
\label{sec:ellipticresults}

The master integrals of the elliptic sector can be computed using the methods developed for elliptic Feynman integrals \cite{Broedel:2017kkb,Adams:2018yfj,Bourjaily:2022bwx}. On the line~\eqref{eq:line}, the master integrals of the sector with denominator $\qG+2y_{12}$ couple to master integrals without any linear denominator, and at each order in $\eps$ the solutions are iterated integrals. Their integration kernels contain rational functions of $\lambda$, the periods~\eqref{eq:periods}, and algebraic functions, and they involve the single elliptic curve~\eqref{eq:quartic}. We do not display these solutions here, since, as we will see, the periods cancel in the wavefunction coefficient.

The master integrals of the three elliptic sectors are therefore genuinely elliptic, and it was from this that ref.~\cite{Benincasa:2024ptf} concluded that the triangle coefficient involves elliptic functions. In the next section, we show that it does not.

%% file: sec4_assembled.tex
\section{The triangle wavefunction coefficient}
\label{sec:assembled}

We now show that the periods of the elliptic curve of section~\ref{sec:elliptic} do not, in fact, appear in the coefficient $\Vc$. We begin by showing that the residue of the integrand of eq.~\eqref{eq:V} on the factors $\qG+2y_e$ is a flat-space amplitude, so that no residue of the integrand involves an elliptic curve (proposition~\ref{prop:residues}). We then derive a differential equation for $\Vc$ on the line~\eqref{eq:line} and show that its operator factors completely into first-order operators. Solving this equation in section~\ref{sec:closedform} gives $\Vc$ in closed form with one leg per site, and in section~\ref{sec:offline} we extend the result to general site energies. Finally, we comment on de~Sitter space in section~\ref{sec:desitter}.

\subsection{Why the periods cancel}
\label{sec:parity}

The key observation is that at the pole $\qG+2y_{12}=0$ the residue of the loop integrand is a flat-space amplitude, whereas the residue of an individual master integral contains the holomorphic differential $\dd\tau/w$ of the curve~\eqref{eq:quartic}. To see this, note that in the recursion~\eqref{eq:recursion3} only the chain in which the edge $y_{12}$ is cut has this pole, which is precisely its total-energy pole. Since the residue of a tree-level coefficient at its total-energy pole is the flat-space amplitude \cite{ArkaniHamed:2017fdk}, the residue of the loop integrand~\eqref{eq:integrand} is
\begin{equation}
\mathop{\rm Res}_{y_{12}=-\qG/2}\psiflat \;=\; \frac{1}{2\,\qG\,\bigl(y_{23}^2-E_2^2\bigr)\bigl(y_{31}^2-E_1^2\bigr)}\,,
\qquad E_1=\tfrac12(X_2+X_3-X_1)\,,\quad E_2=\tfrac12(X_1+X_3-X_2)\,.
\label{eq:residue}
\end{equation}
Equation~\eqref{eq:residue} is simply a product of two Feynman propagators with energies $E_1$ and $E_2$. Crucially, it depends on the edge energies $y_{23}$ and $y_{31}$ only through their squares, which are polynomials in the components of $\vec\ell$, and it is this property that underlies proposition~\ref{prop:residues}. To state the proposition, we write
\begin{equation}
\mu \;=\; \qG\,\psiflat\;y_{12}\,y_{23}\,y_{31}\;\dd y_{12}\,\dd y_{23}\,\dd y_{31}
\label{eq:omega}
\end{equation}
for the rational part of the integrand of eq.~\eqref{eq:V}, which has poles on the nine planes~\eqref{eq:factors}.

\begin{proposition}
\label{prop:residues}
For arbitrary $X_v$ and generic $P_v$, no residue of the integrand of eq.~\eqref{eq:V} on a plane or on a line involves an elliptic curve. More precisely, the following statements hold.
\begin{enumerate}
\item The residue of $\mu$ on the plane $\qG+2y_{12}=0$ is
\begin{equation}
-\frac{\qG}{16}\;\frac{\dd z_{23}\,\dd z_{31}}{\bigl(z_{23}-E_2^2\bigr)\bigl(z_{31}-E_1^2\bigr)}\,,\qquad z_e=y_e^2\,,
\label{eq:residueform}
\end{equation}
and the polynomial $\Bk$ on this plane is of total degree two in $z_{23}$ and $z_{31}$. The same holds on the other two planes $\qG+2y_e=0$.
\item The nine planes meet in pairs along 36 lines. Three of the lines lie at infinity. On 12 lines $\Bk$ is a quadratic polynomial in the coordinate along the line. On the remaining 21 lines $\Bk$ is a quartic polynomial, which for generic $X_v$ has distinct roots, so that $w^2=\Bk$ is an elliptic curve (at $X=P$ most of these quartics become perfect squares; see section~\ref{sec:offline}). Each of these 21 lines lies in one of the planes $\qG+2y_e=0$.
\item On 9 of the 21 lines the residue of $\mu$ vanishes. On each of the other 12 a single edge energy $y$ varies along the line, and $\Bk=b(y^2)$ with $b$ a quadratic polynomial. The residue of $\mu/\sqrt{\Bk}$ is
\begin{equation}
\pm\frac{\qG}{16}\;\frac{\dd Z}{\bigl(Z-E^2\bigr)\sqrt{b(Z)}}\,,\qquad Z=y^2\,,
\label{eq:lineresidue}
\end{equation}
with $E$ one of the energies $E_i$ of the corresponding plane. The same is true if $\Bk^{-1/2}$ is replaced by $\Bk^{-1/2+\eps}$.
\end{enumerate}
\end{proposition}

\noindent The form~\eqref{eq:lineresidue} is a differential on the curve $w^2=b(Z)$, which has genus zero, so the integrand of eq.~\eqref{eq:V} has no residue, on any plane or line, that involves one of the 21 elliptic curves. By contrast, a master integral with a constant numerator and the same two denominators has the residue $\dd y/\sqrt{b(y^2)}$, which is precisely the holomorphic differential of the elliptic curve. The same argument applies to the curve~\eqref{eq:quartic} of section~\ref{sec:elliptic}, which lies at infinity on the plane $\qG+2y_{12}=0$. The form~\eqref{eq:residueform} depends on $y_{23}$ and $y_{31}$ only through $z_{23}$ and $z_{31}$. On the ray $z_{31}=r\,z_{23}$ its product with $\Bk^{-1/2}$ is therefore, at large $z_{23}$, a rational function of $r$ and $z_{23}$ divided by $\sqrt{A(r)}$, which is a differential on a curve of genus zero.

\emph{Proof.} Part~1 is eq.~\eqref{eq:residue} multiplied by the measure, with $y_{23}\,y_{31}\,\dd y_{23}\,\dd y_{31}=\tfrac14\,\dd z_{23}\,\dd z_{31}$, and the statement about $\Bk$ follows from eq.~\eqref{eq:CM}. Part~2 follows from the restriction of eq.~\eqref{eq:CM} to each of the 36 lines. The 12 lines with a quadratic $\Bk$ are those on which two partial energies vanish, and all three edge energies vary along them.

Part~3 follows from part~1. The form~\eqref{eq:residueform} has poles only at $y_{23}=\pm E_2$ and $y_{31}=\pm E_1$. On the plane $\qG+2y_{12}=0$ these are the lines where the energy of site 1, of site 2, or of the pairs of sites $\{2,3\}$ or $\{1,3\}$ vanishes. This gives four lines for each of the three planes. The other planes cross $\qG+2y_{12}=0$ along lines where the form~\eqref{eq:residueform} is regular.\qed

It is instructive to see how the holomorphic differential appears in the individual master integrals and cancels in their sum. The decomposition of the residue~\eqref{eq:residue} into partial fractions is
\begin{equation}
\frac{1}{\bigl(y_{23}^2-E_2^2\bigr)\bigl(y_{31}^2-E_1^2\bigr)} \;=\; \frac{1}{4E_1E_2}\sum_{s,s'=\pm1}\frac{s\,s'}{(y_{23}-s\,E_2)(y_{31}-s'E_1)}\,.
\label{eq:partialfractions}
\end{equation}
Each of the four terms has a non-zero coefficient of the holomorphic differential $\dd\tau/w$ of the curve~\eqref{eq:quartic}. However, the four coefficients are equal in magnitude and carry the signs $s\,s'$, so their sum vanishes. In other words, the periods of the elliptic curve appear in the individual terms into which an integration-by-parts reduction separates $\Vc$, but they cancel between these terms.

The partial energies of the sites require a separate argument, which also illustrates where this simplification ends. At the pole where the partial energy of a single site vanishes, the residue of the loop integrand is
\begin{equation}
\mathop{\rm Res}_{X_1=-y_{31}-y_{12}}\psiflat \;=\; \frac{1}{4\,y_{12}\,y_{31}}\sum_{s,s'=\pm1}s\,s'\;\psi_2\bigl(X_2+s\,y_{12}\,,\,X_3+s'\,y_{31}\,;\,y_{23}\bigr)\,,
\label{eq:triangleresidue}
\end{equation}
with $\psi_2$ given in eq.~\eqref{eq:twosite}. The residue is even in $y_{12}$ and $y_{31}$, the energies of the two edges that end on site~1, but not in $y_{23}$. On the pole, the loop momentum lies on a spheroid with foci at the two points where $y_{12}$ and $y_{31}$ vanish, and $y_{23}$ is the square root of a quadratic function on this surface. Since a single square root of a quadratic can always be rationalized, this residue does not define an elliptic curve. For the polygon with $N$ sites, however, the residue contains $N-2$ such square roots, one for each edge that does not end on the site (figure~\ref{fig:setup}). The box has two, and, as we will see in section~\ref{sec:box}, they do define an elliptic curve.

Since proposition~\ref{prop:residues} holds for arbitrary $X_v$ and generic $P_v$, it also applies with several legs per site, where many more sectors of master integrals contain an elliptic curve (section~\ref{sec:offline}). It does not by itself determine the function space of $\Vc$, however, because the integral over the region $\Bk>0$ is not simply a sum of residues. To determine the function space, we therefore turn to the differential equation.

\subsection{The differential equation}
\label{sec:L0}

We reduce the integrand of eq.~\eqref{eq:V} to master integrals using the integration-by-parts identities of the family~\eqref{eq:family}, as described in appendix~\ref{app:operator}. This expresses $\Vc$ and its $\lambda$-derivatives through the same set of master integrals, and eliminating the master integrals then gives a linear differential operator that annihilates $\Vc$. At generic $\eps$, this operator has order 20. Its limit as $\eps\to0$,
\begin{equation}
\Lzero \;=\; \sum_{i=0}^{20}c_i(\lambda)\,\frac{\dd^i}{\dd\lambda^i}\,,\qquad \Lzero\,\Vc \;=\; 0\,,
\label{eq:L0}
\end{equation}
has polynomial coefficients $c_i$, which we provide in the ancillary files. Its singular points are $\lambda=0,\pm\tfrac13,\pm1$ and $\infty$ (the singular points of $\Ltwo$), together with apparent singularities at which all solutions are analytic.

We emphasize that the limit $\eps\to0$ must be taken after the reduction: since the master integrals have poles at $\eps=0$, terms of order $\eps$ in the reduction contribute to the finite part (appendix~\ref{app:operator}).

\subsection{Complete factorization}
\label{sec:factorization}

The operator $\Lzero$ factors completely over the rational functions of $\lambda$,
\begin{equation}
\Lzero \;=\; c_{20}(\lambda)\,\Bigl(\frac{\dd}{\dd\lambda}-\frac{r_{20}'}{r_{20}}\Bigr)\Bigl(\frac{\dd}{\dd\lambda}-\frac{r_{19}'}{r_{19}}\Bigr)\cdots\Bigl(\frac{\dd}{\dd\lambda}-\frac{r_{1}'}{r_{1}}\Bigr)\,,
\label{eq:factorization}
\end{equation}
where every $r_k$ is a rational function of $\lambda$ and a prime denotes $\dd/\dd\lambda$. We obtain this factorization by right division in the ring of differential operators with coefficients in $\mathbb Q(\lambda)$.

The solutions of a product of first-order operators follow from a single integration for each factor,
\begin{equation}
\Vc(\lambda) \;=\; r_1\int\!\dd\lambda\;\frac{r_2}{r_1}\int\!\dd\lambda\;\frac{r_3}{r_2}\;\cdots\int\!\dd\lambda\;\frac{r_{20}}{r_{19}}\,,
\label{eq:nested}
\end{equation}
with one constant of integration at each step. Since every integrand is a rational function, $\Vc(\lambda)$ is an iterated integral of rational one-forms, and hence a polylogarithmic function. In particular, eq.~\eqref{eq:factorization} contains no irreducible factor of second order, which a period of an elliptic curve would require. On the line~\eqref{eq:line}, this excludes the curve~\eqref{eq:quartic}, its two images under permutations of the sites, and indeed any other elliptic curve. Moreover, the conclusion holds for every right factor of $\Lzero$, so it does not depend on whether $\Lzero$ is the operator of lowest order that annihilates $\Vc$.

\begin{figure}[t]
\centering
\includegraphics[width=0.62\textwidth]{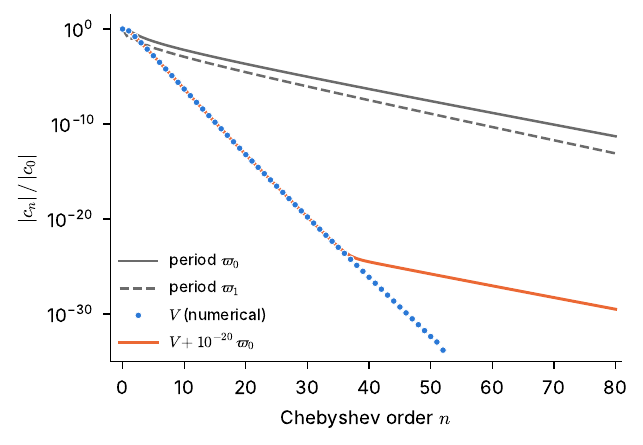}
\caption{Chebyshev coefficients $|c_n|/|c_0|$ of functions on the line~\eqref{eq:line}, expanded on $0.345\le\lambda\le0.985$. The elliptic periods $\varpi_0$ and $\varpi_1$ of eq.~\eqref{eq:periods} (gray) have logarithmic singularities at $\lambda=\tfrac13$ and $\lambda=1$, respectively, just outside this interval, so their coefficients decay slowly. The coefficients of the numerical values of $\Vc$ (blue) decay much faster; they are shown up to the order at which they reach the precision of the numerical values, about $10^{-34}$. If $\Vc$ contained even a small multiple of a period, its coefficients would follow the slow decay at large $n$, as for $\Vc+10^{-20}\varpi_0$ (orange). No such component is present.}
\label{fig:cheb}
\end{figure}

The numerical values of $\Vc$ are consistent with this result. While the solutions of $\Ltwo$ have logarithmic singularities at $\lambda=\tfrac13$ or at $\lambda=1$, the Chebyshev coefficients of $\Vc$ decay as those of a function that is analytic at both points (figure~\ref{fig:cheb}), as is the closed form of section~\ref{sec:closedform}.

\subsection{The closed form}
\label{sec:closedform}

On the line~\eqref{eq:line}, the solutions of $\Lzero$ turn out to be polylogarithms of weight at most two. We solve $\Lzero\,g=0$ exactly using an ansatz in multiple polylogarithms $G(l_1,\ldots,l_w;\lambda)$ with letters $l_i\in\{0,\pm1,\pm\tfrac13\}$ and coefficients that are rational functions of $\lambda$. We find 5 solutions of weight zero, 19 of weight at most one, and 20 of weight at most two, with no further solutions at weights three and four. The coefficient $\Vc$ is then a combination of the 20 solutions with constant coefficients, which are fixed by analyticity at $\lambda=1$ together with the numerical values of $\Vc$ (appendix~\ref{app:line}).

Written in terms of the site energies, the result on the line suggests the following formula for one leg per site and arbitrary momenta:
\begin{equation}
\Vc \;=\; \frac{\sqrt2\,\pi}{8}\left[\frac{\sum_i X_i\,\mathrm{Li}_2(u_i)}{\xi_1\xi_2\xi_3}+\frac{8\,X_1X_2X_3}{\qG\,\xi_1^2\xi_2^2\xi_3^2}\;\mathcal W\right],
\label{eq:closedform}
\end{equation}
\begin{equation}
\mathcal W \;=\; \sum_i X_i\xi_i\,\mathrm{Li}_2(u_i)+2\sum_{i<j}X_iX_j\,\ln(1-u_i)\ln(1-u_j)-\frac{\pi^2}{12}\sum_{i<j}\xi_i\xi_j\,,
\qquad \xi_i=\qG-2X_i\,,\quad u_i=\frac{\xi_i}{\qG}\,.
\label{eq:closedformW}
\end{equation}
On the line~\eqref{eq:line}, the function space of eq.~\eqref{eq:closedform} is exact, while its constants are determined numerically. For other momenta, eq.~\eqref{eq:closedform} is an ansatz with 14 rational coefficients. The result on the line determines these 14 coefficients uniquely, and the formula then agrees with numerical values of the loop integral for other triangles of momenta (appendix~\ref{app:general}). Equation~\eqref{eq:closedform} can also be derived directly from the integral~\eqref{eq:V}, without the differential equation or numerical values, by carrying out the three integrations over the edge energies, as we sketch in appendix~\ref{app:derivation}.

The variables $\xi_i$ are positive by the triangle inequalities, with $\xi_i=0$ corresponding to a folded configuration. The expression remains finite there, since the bracket vanishes to second order, and it is also finite when one of the momenta becomes soft. Interestingly, the combination $\qG^2\,\Vc$ is nearly independent of the shape of the triangle of momenta: it varies by $5.4\%$ over all triangles, from its minimum for the equilateral triangle to its maximum in the soft limits.

With one leg per site, the one-loop triangle coefficient is therefore a function of transcendental weight two built from three dilogarithms (here we count the weight of the bracket in eq.~\eqref{eq:closedform}, not its overall factor of $\pi$), the same weight as a one-loop amplitude in flat space. To our knowledge, the only earlier closed form for this coefficient is that of ref.~\cite{Salcedo:2022aal}, for the configuration in which one external momentum vanishes at arbitrary site energies. In that configuration the integration region reduces to that of the two-site graph, and the result is again a combination of dilogarithms.

\subsection{General site energies}
\label{sec:offline}

As discussed in section~\ref{sec:kinematics}, the integrals of eq.~\eqref{eq:siteintegrals} require $\Vc$ at $X_v>P_v$. Here we first state the result and define its terms, and then describe how we obtained it from the master integrals on a second line, the symbol, and numerical values of the loop integral. For these site energies, the coefficient is again a polylogarithmic function of weight two, now with letters that contain square roots,
\begin{equation}
\Vc(X;P) \;=\; \frac{\sqrt2\,\pi}{8}\,\mathrm{Re}\biggl[\,-\frac14\,\frac{\qG}{\sqrt{R_0}}\,F_0-\frac12\sum_{e}\frac{\qG}{\sqrt{R_e}}\,F_e
-\sum_{i<j}\frac{\rho^-_{ij}}{\sqrt{T^-_{ij}}}\,F^-_{ij}+\sum_{i<j}\frac{\rho^+_{ij}}{\sqrt{T^+_{ij}}}\,F^+_{ij}\biggr]\,.
\label{eq:closedformgeneral}
\end{equation}
Here $R_0$ is $-2\Bk$ evaluated at the point where the three partial energies of the sites vanish, while the three $R_e$ are $-2\Bk$ evaluated at the points where $\qG+2y_e$ and the partial energies of the two sites at the ends of $e$ vanish. The remaining radicands and prefactors are
\begin{equation}
\begin{aligned}
T^\pm_{ij} &= \kappa(P)+4P_i^2X_j^2+4P_j^2X_i^2\pm4X_iX_j\bigl(P_i^2+P_j^2-P_k^2\bigr)\,,\\
\rho^+_{ij}&=\frac{X_k}{X_i+X_j-X_k}\,,\qquad \rho^-_{ij}=\frac{X_k\,\qG}{(X_i-X_j)^2-X_k^2}\,,
\end{aligned}
\label{eq:Tpm}
\end{equation}
where $\kappa(P)$ is the K\"all\'en function of the $P_v^2$ and $k$ is the third site. Each $F$ is a combination of dilogarithms of the form
\begin{equation}
\mathcal D(x,\bar x) \;=\; \mathrm{Li}_2(x)-\mathrm{Li}_2(\bar x)+\tfrac12\ln(x\bar x)\,\ln\frac{1-x}{1-\bar x}\,,
\label{eq:BW}
\end{equation}
where $\bar x$ is obtained from $x$ by changing the sign of the square root. For a positive radicand, $x$ and $\bar x$ are real, and we take the real parts of the dilogarithms and the absolute values of the arguments of the logarithms. For a negative radicand, $\bar x$ is the complex conjugate of $x$, and $\mathcal D$ is $2\mathrm i$ times the Bloch--Wigner function. With these definitions, each term in the bracket of eq.~\eqref{eq:closedformgeneral} is real. Since permutations of the sites relate the three $F_e$, the three $F^+_{ij}$, and the three $F^-_{ij}$ to each other, there are four independent functions in total. Notably, there is no term without a square root and no additive constant.

The function $F_0$ has a particularly nice geometric form. At the point where the three partial energies of the sites vanish, the $P_v$ and the values of $-y_e$ are the six edge lengths of a flat tetrahedron (where a negative value is understood as a negative length). Letting $f_k$ denote the perimeters of its four faces and $c_j$ the perimeters of the three closed paths through all four vertices,
\begin{equation}
\mathcal P(h)\;=\;\prod_{k=1}^{4}(h-f_k)\;-\;h\prod_{j=1}^{3}(h-c_j)
\label{eq:tetraquadratic}
\end{equation}
is a quadratic polynomial in the auxiliary variable $h$, whose discriminant is the radicand $R_0$. With $h_\pm$ the roots of $\mathcal P$ and $\mathcal D_a=\mathcal D(1-a/h_-,1-a/h_+)$, we then have
\begin{equation}
F_0\;=\;-2\,\Bigl[\,\sum_{k=1}^{4}\mathcal D_{f_k}-\sum_{j=1}^{3}\mathcal D_{c_j}\Bigr]\,.
\label{eq:F0}
\end{equation}
Strikingly, when $R_0<0$ each term of eq.~\eqref{eq:F0} is proportional to the volume of an ideal hyperbolic tetrahedron. The function $F_e$ is a sum of nine terms of the same type, while $F^+_{ij}$ and $F^-_{ij}$ are sums of five and four terms, respectively; we give them in appendix~\ref{app:functions}.

We now turn to how eq.~\eqref{eq:closedformgeneral} was obtained. On the second line, $X=P+\tfrac13(1,1,1)$ with $P=(2\lambda,\lambda,1)$, there are 134 master integrals, and fifteen sectors in which two of the factors~\eqref{eq:factors} vanish together have an irreducible homogeneous equation of second order. Their operators are equivalent to the Picard--Fuchs operators of the elliptic curves that $\Bk$ defines on the corresponding lines of proposition~\ref{prop:residues}. At $X=P$, most of these curves degenerate, since $\Bk$ becomes a perfect square there.

Once again, the periods of these curves do not enter the coefficient: on this line, its numerical values are reproduced by a sum of ten terms, each a polylogarithm of weight two divided by a square root. The ten square roots are those of eq.~\eqref{eq:closedformgeneral}, and all ten become perfect squares at $X=P$.

To move to general kinematics, we use the symbol \cite{Goncharov:2010jf}. Recall that the symbol of a function of weight two is a sum of ordered pairs of letters, in which the first entry of each pair determines where the function has branch points; moreover, a sum of pairs is the symbol of a function only if it satisfies an integrability condition. We consider symbols whose first entries are the positive sums of energies that occur on the second line, with $i$, $j$, and $k$ distinct,
\begin{equation}
\qG\,,\quad X_i+P_i\,,\quad X_i+X_j+P_k\,,\quad X_i+P_j+P_k\,,\quad P_1+P_2+P_3\,.
\label{eq:firstentries}
\end{equation}
The second entries are letters of the form $(p+\sqrt R)/(p-\sqrt R)$, where $R$ is one of the ten radicands and $p$ is a polynomial for which $p^2-R$ is a product of linear functions of the energies, so that each symbol is odd under the change of sign of its root. For each of the ten square roots, there is exactly one integrable symbol of this kind, and the prefactors $\qG/\sqrt{R}$ and $\rho^\pm_{ij}/\sqrt{T^\pm_{ij}}$ are the corresponding leading singularities. The coefficient is therefore fixed up to ten rational numbers. Fitting these ten numbers to numerical values of the loop integral at generic points gives the coefficients $-\tfrac14$, $-\tfrac12$, $-1$, and $+1$ of eq.~\eqref{eq:closedformgeneral}. These numbers also follow from the integrand itself: the part of the symbol whose first entry is $X_1+P_1$ can be computed directly from the integral~\eqref{eq:V}, and for each of the seven square roots that involve site~1 it agrees with eq.~\eqref{eq:closedformgeneral}. The coefficients are residues of the integrand at the points and on the lines where its partial energies vanish together (appendix~\ref{app:coefficients}). With these values, eq.~\eqref{eq:closedformgeneral} agrees with numerical values of the loop integral throughout $X_v\ge P_v$ (appendix~\ref{app:general}). As a further check, at $X=P$ the ten square roots become rational, and the symbol of eq.~\eqref{eq:closedformgeneral} reduces to that of eq.~\eqref{eq:closedform}.

\begin{figure}[t]
\centering
\includegraphics[width=0.62\textwidth]{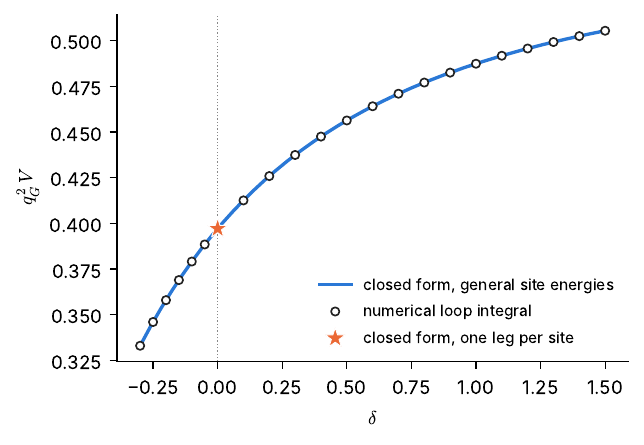}
\caption{The combination $\qG^2\,\Vc$ for the triangle of momenta $P=(\tfrac75,\tfrac7{10},1)$, with the energy of every site shifted by the same amount, $X_v=P_v+\delta$. The line is eq.~\eqref{eq:closedformgeneral}, the points are numerical values of the loop integral, and the star is the formula for one leg per site, eq.~\eqref{eq:closedform}, at $\delta=0$. For $\delta<0$, where $X_v<P_v$, the line is the analytic continuation of section~\ref{sec:offline}. At $\delta=0$ (dotted) all ten square roots of eq.~\eqref{eq:closedformgeneral} become rational; individual terms are singular there, but their sum is smooth. The formula and the numerical values agree to 11 digits, the precision of these points (appendix~\ref{app:operator}).}
\label{fig:delta}
\end{figure}

With the definition~\eqref{eq:BW}, eq.~\eqref{eq:closedformgeneral} holds for $X_v\ge P_v>0$. Individual terms are singular where the prefactors~\eqref{eq:Tpm} have poles, where a radicand vanishes, where $P_k=|X_i-X_j|$, and where the leading coefficient of $\mathcal P$ vanishes for the primed lengths of appendix~\ref{app:functions}, but the sum has a finite limit at each of these points, and at $X=P$ this limit is eq.~\eqref{eq:closedform}. Although we have so far focused on $X_v\ge P_v$, the coefficient is defined for all $X_v>0$, and its integrand has no singularity in that region. For $X_v<P_v$ and a positive radicand, each term of eq.~\eqref{eq:closedformgeneral} differs from its value with the definition~\eqref{eq:BW} by an integer multiple of $\pi^2/2$, with the integer determined by analytic continuation along the ray $P\to tP$ from $t=0$. The continued expression is smooth across $X_v=P_v$ (figure~\ref{fig:delta}). The ancillary files contain a numerical implementation of eq.~\eqref{eq:closedformgeneral} with these branch assignments, including the continuation to $X_v<P_v$.

To summarize, the one-loop triangle coefficient is a polylogarithmic function of weight two. This statement is exact on the line~\eqref{eq:line}, where the operator $\Lzero$ factors completely, and for other kinematics it rests on the agreement of eq.~\eqref{eq:closedformgeneral} with numerical values of the loop integral. The elliptic curves of the master integrals (three at $X=P$, and many more for $X_v>P_v$) do not enter, while the ten square roots do, appearing in the letters of the coefficient for $X_v>P_v$.

\subsection{De~Sitter space}
\label{sec:desitter}

For a cubic interaction in de~Sitter space, the integrals~\eqref{eq:siteintegrals} have $\sigma=0$, and they change the behavior of the loop integrand at large loop momentum from $|\vec\ell|^{-5}$ to
\begin{equation}
\int_0^\infty\!\dd^3\omega\;\psiflat(X+\omega;y) \;\longrightarrow\; \frac{\zeta(3)}{8\,|\vec\ell|^{3}}\,,
\label{eq:dslarge}
\end{equation}
so the loop integral of the de~Sitter coefficient diverges logarithmically. With a cutoff $\Lambda$ on the comoving loop momentum, the coefficient of $\ln\Lambda$ is $\pi\,\zeta(3)/2$. With one leg per site, symmetry then determines the dependence on the momenta: at late times, the coefficient is a three-point function of operators with $\Delta=2$ in a three-dimensional conformal field theory \cite{Bzowski:2013sza}. Assuming that the cutoff breaks the symmetry only through the logarithm, we have
\begin{equation}
\psi_3^{\rm dS} \;=\; \int^{\Lambda}\!\dd^3\ell\int_0^\infty\!\dd^3\omega\;\psiflat(X+\omega;y) \;=\; -\frac{\pi\,\zeta(3)}{2}\,\ln\frac{P_1+P_2+P_3}{\Lambda}+b_0\,,
\label{eq:dsform}
\end{equation}
where $b_0$ is a number that depends on the definition of the cutoff. A numerical evaluation at two kinematic points agrees with this form, although we do not derive eq.~\eqref{eq:dsform} from eq.~\eqref{eq:closedformgeneral}. For the corresponding correlator, the loop integral converges, and symmetry allows only a constant, whose numerical value is $-3\pi^2\zeta(3)$.

With one leg per site, the de~Sitter triangle is therefore determined by just two numbers (the coefficient of the logarithm and $b_0$), regardless of the function space of $\Vc$. The function space of $\Vc$ becomes important when several legs meet at a site, or when the external legs break the de~Sitter boosts. In those cases, the de~Sitter and power-law coefficients are integrals of eq.~\eqref{eq:closedformgeneral} over the site energies.

%% file: sec5_correlators.tex
\section{Correlators}
\label{sec:correlators}

Having established that the elliptic periods cancel in the wavefunction coefficient, we now turn to the equal-time correlator. In this section, we show that the integrand of the triangle correlator has no pole on the planes $\qG+2y_e=0$, which are precisely the planes of the elliptic sectors of section~\ref{sec:elliptic}. The absence of the factors $\qG+2y_e$ from correlators is already known \cite{AguiSalcedo:2023nds,Chowdhury:2023arc}, and the relation between correlators and wavefunction coefficients has been studied from several directions \cite{Chowdhury:2025ohm,Chowdhury:2026upp,Chowdhury:2026dwm}. In particular, ref.~\cite{Benincasa:2024ptf} noted this absence in connection with the elliptic sector and asked whether the correlator requires elliptic functions at all. Here we derive the cancellation from the recursion relations and then consider the other singular loci of the correlator integrand. Throughout this section, all statements concern integrands at fixed loop momentum.

\subsection{The correlator integrand}
\label{sec:bornrule}

The Born rule~\eqref{eq:born} expresses a correlator through products of wavefunction coefficients. At one loop, the integral in eq.~\eqref{eq:born} also includes the profile of the field running around the loop; with the Gaussian weight $|\Psi|^2$, this integral joins two external legs of energy $y_e$ and gives a factor $1/(2y_e)$. The correlator integrand of a graph $G$ is therefore a sum over the sets of edges that are cut into such pairs of legs. Writing $\hat\psi_G=\prod_e(2y_e)\,\psi_G$, for vertex factors that do not depend on time the expansion of the Born rule at one loop (section~3 of ref.~\cite{AguiSalcedo:2023nds} and appendix~A of ref.~\cite{Chowdhury:2023arc}) gives
\begin{equation}
C_G \;=\; \sum_{S\,\subseteq\,\text{edges}} 2^{\,n(S)-1}\prod_{j=1}^{n(S)}\hat\psi_{H_j}\,,
\label{eq:KG}
\end{equation}
where the graph with the edges in $S$ cut has $n(S)$ connected components $H_j$, and each cut edge adds its energy $y_e$ to the two sites at its ends. The correlator integrand is then $C_G$ divided by $\prod_e 2y_e$ and by factors that depend only on the external legs. Note that the first terms of eq.~\eqref{eq:KG}, namely the loop and the chains in which one edge is cut, appear with the same coefficient.

\subsection{Cancellation of the factors \texorpdfstring{$\qG+2y_e$}{q\_G + 2 y\_e}}
\label{sec:cancellation}

Consider the polygon with $N$ sites, and let $G_e$ denote the chain in which the edge $e$ is cut. The only terms of eq.~\eqref{eq:KG} with poles at $\qG+2y_e=0$ are the loop and the chains, which appear in the combination
\begin{equation}
\Phi_G \;=\; \hat\psi_G+\sum_{e}\hat\psi_{G_e}\,.
\label{eq:FG}
\end{equation}
The recursion of ref.~\cite{ArkaniHamed:2017fdk}, which is eq.~\eqref{eq:recursion3} for the triangle, gives
\begin{equation}
\qG\,\hat\psi_G \;=\; \sum_e 2y_e\,\hat\psi_{G_e}\,,\qquad
(\qG+2y_e)\,\hat\psi_{G_e} \;=\; \sum_{f\ne e}2y_f\,\hat\psi_{G_e\setminus f}\,,
\label{eq:tworecursions}
\end{equation}
where $G_e\setminus f$ is the pair of chains in which a second edge $f$ is cut, and the second relation holds because the total energy of the chain $G_e$ is $\qG+2y_e$. Adding the first relation to the sum of the second over $e$, the terms $2y_e\hat\psi_{G_e}$ cancel, and we find
\begin{equation}
\qG\,\Phi_G \;=\; \sum_{e\ne f}2y_f\,\hat\psi_{G_e\setminus f}\,.
\label{eq:FGidentity}
\end{equation}
On the right-hand side, $\hat\psi_{G_e\setminus f}$ is the product of the coefficients of the two chains. Each chain contains only one end of $e$ and one end of $f$, so no connected set of its sites contains the energy of a cut edge twice, and all poles are partial energies of $G$. The poles of $\hat\psi_G$ and of $\hat\psi_{G_e}$ at $\qG+2y_e=0$ therefore cancel in $\Phi_G$. The remaining terms of eq.~\eqref{eq:KG} have two or more edges cut, and their poles are likewise partial energies of $G$. For the triangle, the denominator of $\Phi_G$ thus consists of $\qG$ and the six partial energies of eq.~\eqref{eq:factors}. The argument extends by induction to polygons with trees attached, and ref.~\cite{AguiSalcedo:2023nds} derives the cancellation more generally, for any number of loops and any time dependence of the background, as a cosmological analog of the Kinoshita--Lee--Nauenberg theorem.

Physically, the factors $\qG+2y_e$ originate from the boundary condition of the wavefunction. The propagator~\eqref{eq:propagator} is the sum of a time-ordered term and a factorized term, $e^{-y(t+t')}$, the latter of which makes the propagator vanish at the late-time boundary. A factor containing $y_e$ twice requires both ends of the edge to add the energy $+y_e$ to the same set of sites. Since the time-ordered term adds $+y_e$ only at its earlier end, the factors $\qG+2y_e$ must come from the factorized term. In the correlator, the integral over the profile produces the chains $G_e$, in which the edge is replaced by the same pair of external legs, and the two contributions cancel.

\subsection{Other singular loci of the correlator integrand}
\label{sec:census}

The cancellation removes the curve of section~\ref{sec:elliptic} from the correlator integrand of the triangle, and it turns out that the other singular loci do not contain elliptic curves either. By proposition~\ref{prop:residues}, every line on which $\Bk$ defines an elliptic curve lies in one of the planes $\qG+2y_e=0$. Since the correlator integrand has no pole on these planes, it has no residue on any of these lines. The integrand of eq.~\eqref{eq:V}, on the other hand, has residues on 12 of them, but by proposition~\ref{prop:residues} these residues do not involve the periods of an elliptic curve. For the triangle, then, neither integrand has a residue involving an elliptic curve, though for different reasons: the correlator integrand has no pole on the lines with an elliptic curve, while the residues of the wavefunction integrand on these lines are differentials on curves of genus zero.

For the box and the pentagon, the result depends on the dimension. In $d=N$, the $N$ edge energies are independent coordinates on the space of loop momenta, and the measure is the inverse square root of the Cayley--Menger polynomial of the simplex with $N+1$ vertices (for the triangle, this is the physical dimension). In $d=N$, the integrands of the wavefunction coefficients of the box and the pentagon have residues on lines with an elliptic curve, including lines on which only partial energies vanish, whereas their correlator integrands have none. We have not studied singular loci of dimension two or higher. In $d=3$, however, the exponent of the Cayley--Menger polynomial of the box is an integer, and the curves are of a different kind; as we show in section~\ref{sec:box}, the correlator of the box in $d=3$ does involve elliptic curves.

Finally, the cancellation also holds for time-dependent vertex factors and at two loops. With vertex factors $(-\eta)^{-(1+\sigma)}$, the coefficients in eq.~\eqref{eq:KG} contain a phase for each connected component that depends on its number of sites. Since the loop and the chains $G_e$ have the same sites, $\Phi_G$ is unchanged, and the integrals of eq.~\eqref{eq:siteintegrals} apply to $\Phi_G$ as a whole. In the sunrise, the kite, and five other two-loop graphs with up to four sites, every factor that contains an edge energy more than once is absent from the correlator integrand, in agreement with ref.~\cite{AguiSalcedo:2023nds}. Whether the integrand of a two-loop correlator has residues that involve an elliptic curve remains an open question.

%% file: sec6_box.tex
\section{The box}
\label{sec:box}

Unlike the triangle, the one-loop box is not polylogarithmic. In this section, we show that the residue of its loop integrand at the partial-energy pole of a single site has two square roots, which together define an elliptic curve. We give the discontinuity of the box at threshold in closed form in terms of elliptic integrals, describe the K3 surface that appears above the threshold, and compute the box and its correlator on a line. In the last subsection, we summarize what we know for general momenta. We use the notation of section~\ref{sec:setup}, with sites $v=1,\ldots,4$ and edge energies $y_{12}$, $y_{23}$, $y_{34}$, and $y_{41}$ (figure~\ref{fig:setup}(b)), and we take the wavefunction coefficient to be $V_4=\int\dd^3\ell\;\psi_4$, without a prefactor.

\subsection{The partial-energy pole of a single site}
\label{sec:boxresidue}

For the box, the residue on the factors $\qG+2y_e$ is once again a flat-space amplitude, depending on the other edge energies only through their squares, and the residues on the partial energies of two or three sites do not define an elliptic curve either. Instead, the elliptic curve comes from the partial energy of a single site. At the pole $X_1+y_{41}+y_{12}=0$, the residue of the loop integrand is
\begin{equation}
\mathop{\rm Res}_{X_1=-y_{41}-y_{12}}\psi_4 \;=\; \frac{1}{4\,y_{12}\,y_{41}}\sum_{s,s'=\pm1}s\,s'\;\psi_{\rm chain}\bigl(X_2+s\,y_{12}\,,\;X_3\,,\;X_4+s'\,y_{41}\,;\;y_{23}\,,\,y_{34}\bigr)\,,
\label{eq:boxresidue}
\end{equation}
where $\psi_{\rm chain}$ is the coefficient~\eqref{eq:chain} of the chain formed by the three other sites, in analogy with eq.~\eqref{eq:triangleresidue}. The residue is even in $y_{12}$ and $y_{41}$, but not in $y_{23}$ or $y_{34}$, the energies of the two edges that do not end on site~1. The same formula holds for every polygon, with the chain of the other sites in place of eq.~\eqref{eq:chain}.

At the pole, the loop momentum lies on a spheroid with foci at the two points where $y_{12}$ and $y_{41}$ vanish, and each of the other edge energies is the square root of a quadratic function on this surface. This is where the box and the triangle part ways. The triangle has only one such edge, and a single square root of a quadratic can always be rationalized (section~\ref{sec:parity}); the box, however, has two. Along a line on the spheroid, the two square roots define the curve $Y^2=z_{23}\,z_{34}$, where $z_e=y_e^2$ is quadratic in the coordinate along the line, and this curve has genus one. For the pentagon, the three square roots have six branch points and define a curve of genus five.

\subsection{The threshold discontinuity}
\label{sec:boxoneleg}

With one leg per site ($X_v=P_v$), the discontinuity of $V_4$ at its threshold reduces to a one-dimensional integral. More precisely, the discontinuity of $V_4$ across the branch cut that starts at $X_1=-P_1$ is the integral of the residue~\eqref{eq:boxresidue} over the spheroid $y_{41}+y_{12}=-X_1$, and at the threshold the spheroid degenerates to the segment between its foci. Parametrizing the segment by $\chi=y_{41}-y_{12}$ with $-P_1\le \chi\le P_1$, we have
\begin{equation}
y_{23}^2 \;=\; \frac{(P_1+\chi)\,P_2^2+(P_1-\chi)\,P_{12}^2}{2P_1}-\frac{P_1^2-\chi^2}{4}\,,\qquad
y_{34}^2 \;=\; \frac{(P_1+\chi)\,P_{41}^2+(P_1-\chi)\,P_4^2}{2P_1}-\frac{P_1^2-\chi^2}{4}\,,
\label{eq:segment}
\end{equation}
on the segment, with $P_{12}=|\vec P_1+\vec P_2|$ and $P_{41}=|\vec P_4+\vec P_1|$. We define the threshold discontinuity as
\begin{equation}
D_0 \;=\; \frac{\pi}{4P_1}\int_{-P_1}^{P_1}\!\dd\chi\;\sum_{s,s'=\pm1}s\,s'\;\psi_{\rm chain}\Bigl(P_2+s\,\tfrac{P_1-\chi}{2}\,,\;P_3\,,\;P_4+s'\,\tfrac{P_1+\chi}{2}\,;\;y_{23}\,,\,y_{34}\Bigr)\,.
\label{eq:D0}
\end{equation}
For generic momenta, the four zeros of $z_{23}\,z_{34}$ are distinct and form two complex-conjugate pairs, and eq.~\eqref{eq:D0} is an integral on an elliptic curve whose $j$-invariant depends on the shape of the quadrilateral of momenta.

Importantly, the threshold discontinuity is not merely a feature of the integrand, but part of the wavefunction coefficient itself. When the momentum of the leg at site~1 becomes soft, the coefficient with one leg per site takes the form
\begin{equation}
V_4 \;=\; -D_0\,\ln 2P_1 \;+\; \text{terms analytic at } P_1=0\,,
\label{eq:softlimit}
\end{equation}
and numerical values of $V_4$ at two quadrilaterals that do not lie in a plane agree with eq.~\eqref{eq:softlimit}.

\begin{figure}[t]
\centering
\includegraphics[width=0.62\textwidth]{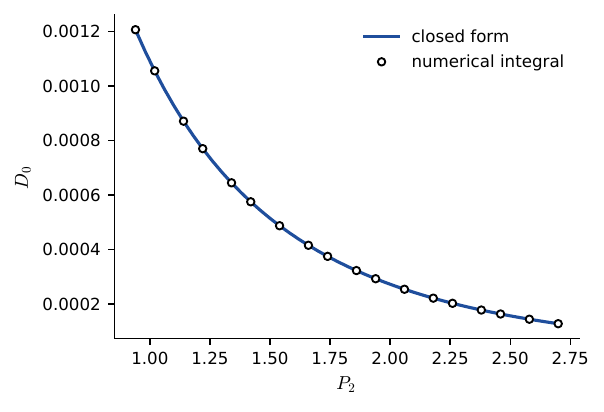}
\caption{The threshold discontinuity $D_0$ of the box with one leg per site, for $P_1=1$, $P_3=2$, $P_4=\tfrac74$, $P_{12}=\tfrac95$ and $P_{41}=2$, as a function of $P_2$. Here $P_{12}$ and $P_{41}$ are the lengths of the two diagonals of the quadrilateral. The line is the closed form~\eqref{eq:D0closed} and the points are numerical values of the integral~\eqref{eq:D0}.}
\label{fig:boxD0}
\end{figure}

The integral~\eqref{eq:D0} can be evaluated in closed form. Its integrand is a rational function of $\chi$, $y_{23}$, and $y_{34}$ with simple poles, and partial fractions reduce it to logarithms and to the integrals of the first and third kind on the curve $Y^2=z_{23}\,z_{34}$ of section~\ref{sec:boxresidue}. The result is
\begin{equation}
D_0 \;=\; \frac{\pi}{4P_1}\Bigl[\;\frac14\sum_{\rm poles}\rho\;\bigl(L_0+\eta\,L_{23}+\eta'L_{34}+\eta\,\eta'\,L_{\Pi}\bigr)\;+\;4\,p_0\,J_F\Bigr]\,,
\qquad J_F=\int_{-P_1}^{P_1}\frac{\dd\chi}{4\,Y}=2R_F\,,
\label{eq:D0closed}
\end{equation}
where $L_0$, $L_{23}$, and $L_{34}$ are logarithms, and $L_\Pi$ contains a logarithm and the symmetric elliptic integrals $R_F$ and $R_J$ of Carlson \cite{Carlson:1995}. The sum runs over the poles of the integrand on the four sheets $(\eta,\eta')=(\pm y_{23},\pm y_{34})$ of the two roots, and the coefficients $\rho$ and $p_0$ are algebraic functions of the six lengths in eqs.~\eqref{eq:segment} and~\eqref{eq:D0}; we derive the complete expressions in appendix~\ref{app:D0}. That these elliptic integrals do not combine into logarithms is shown in section~\ref{sec:boxmonodromy}. Figure~\ref{fig:boxD0} compares eq.~\eqref{eq:D0closed} with numerical values of the integral~\eqref{eq:D0}.

For momenta in a plane, one of the poles becomes a double pole, and eq.~\eqref{eq:D0closed} then contains the additional term of eq.~\eqref{eq:doublepole}. With this term, the formula holds for planar quadrilaterals in general position. In the planar case all poles are real, and the result does not depend on the branches of the logarithms. For momenta that do not lie in a plane, the branches at the complex poles follow from an analytic continuation, which agrees with numerical values but which we have not proved.

\subsection{The K3 surface above the threshold}
\label{sec:boxk3}

Above the threshold, the discontinuity becomes an integral over the double cover of the spheroid defined by the two roots, which is a K3 surface fibered by elliptic curves. Its periods obey a differential equation in $y_{41}+y_{12}$, whose order is 5 when the momenta form a regular tetrahedron and 8 for generic momenta. For the regular tetrahedron, we have the operator exactly (eq.~\eqref{eq:L5} in appendix~\ref{app:L5}). It is not a symmetric power of a second-order operator, so its solutions are not simply products of periods of an elliptic curve. Instead, its five solutions obey the quadratic relation~\eqref{eq:quadrel}, a characteristic property of the periods of a K3 surface.

The loop integrand has a component along these periods. With several legs per site, the leading term of the residue at large site energies generates all five solutions of the fifth-order operator, so the periods of the K3 surface enter the loop integrand of the box. With one leg per site, the evidence is as follows. The solutions of the fifth-order operator have square-root branch points at complex values of $y_{41}+y_{12}$, and the discontinuity of $V_4$ has the same branch points with the predicted coefficient. Moreover, an exact reduction of the loop integrand, for the regular tetrahedron and with three site energies at one leg per site, gives differential equations whose solutions include the period of the holomorphic form of the K3 surface, and the same holds for the integrand of the correlator. With one leg per site, we have not examined generic momenta.

\subsection{The box on a line}
\label{sec:boxline}

Just as for the triangle in section~\ref{sec:assembled}, we can obtain exact differential equations for the box on a line in kinematic space. We take the momenta to form a regular tetrahedron of unit side, with site energies
\begin{equation}
X_v \;=\; X \;=\; 1+t\,,\qquad v=1,\ldots,4\,,
\label{eq:boxlinedef}
\end{equation}
so that $t=0$ corresponds to the case of one leg per site. We denote by $C_4(t)$ the correlator corresponding to $V_4$, integrated over the loop momentum and normalized so that its leading term at large $X$ is $X^{-4}\int\dd^3\ell/(y_{12}y_{23}y_{34}y_{41})$, eq.~\eqref{eq:boxlargeX}. An integration-by-parts reduction with the regulator of eq.~\eqref{eq:V} gives exact first-order differential systems in $t$ for both $V_4$ and $C_4$. Taking the limit $\eps\to0$ requires the double poles of the master integrals in $\eps$, whose coefficients are residues at infinite loop momentum and do not depend on the kinematics.

These systems determine the function space on the line. An exact rational change of basis brings the system for $C_4$ to a form with simple poles at the zeros of 19 polynomials in $X$ of degree at most four (listed in eq.~\eqref{eq:letters}), and the system has no other singular points at finite $X$. Its irreducible parts are the fifth-order operator of section~\ref{sec:boxk3}, the second-order operators of three elliptic curves, and first-order operators with rational and square-root letters. The three curves are those on which two partial energies vanish, while the curve of section~\ref{sec:boxoneleg} degenerates on this line and does not appear. The periods of the three curves are hypergeometric functions of their $j$-invariants (eq.~\eqref{eq:period}), and they enter both $V_4$ and $C_4$. The monodromy of the solutions indicates that the solutions of the fifth-order operator enter as well, although we have not separated them from the solutions of the first-order operators. The solutions of the systems agree with numerical values of both functions and continue smoothly to $t=0$ (appendix~\ref{app:uv}). The ancillary files contain the two systems along with a program that evaluates their solutions.

\subsection{The correlator on the line}
\label{sec:boxclosed}

\begin{figure}[t]
\centering
\includegraphics[width=0.62\textwidth]{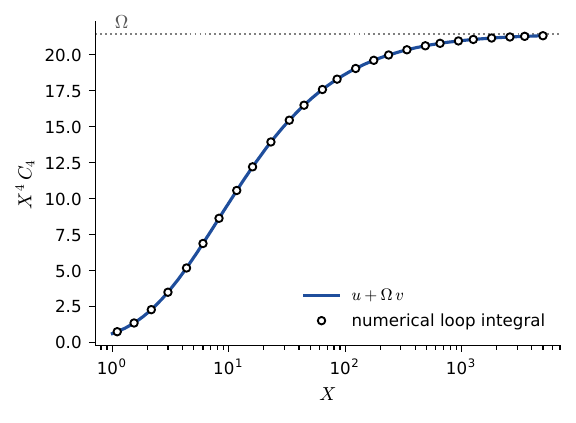}
\caption{The correlator of the box on the line of section~\ref{sec:boxline}, multiplied by $X^4$. The line is eq.~\eqref{eq:boxclosed} and the points are numerical values of the loop integral. The dotted line is the constant $\Omega$, which is the limit at large $X$.}
\label{fig:boxline}
\end{figure}

Remarkably, on this line the differential system determines the correlator completely. The result is
\begin{equation}
\begin{aligned}
C_4(t)\;&=\;u(t)+\Omega\,v(t)\,,\\
\Omega\;&=\;\int\!\dd^3\ell\;\frac{1}{y_{12}\,y_{23}\,y_{34}\,y_{41}}\;=\;21.4444295139\ldots\,,
\end{aligned}
\label{eq:boxclosed}
\end{equation}
where $u$ and $v$ are two solutions of the system for $C_4$, defined in appendix~\ref{app:uv}, and in the integral the four edge energies are the distances from $\vec\ell$ to the vertices of the regular tetrahedron. We show the result in figure~\ref{fig:boxline}.

Notably, the functions $u$ and $v$ follow from conditions that do not use any value of $C_4$. Since the loop integral converges for $t>-1$, $C_4$ must be single-valued around the singular points of the system in that range, and the solutions that are single-valued around the six points listed in appendix~\ref{app:uv} form a space of dimension 4. At large $X$, the correlator has the expansion
\begin{equation}
C_4\;=\;\sum_{n\ge4}X^{-n}\bigl(\alpha_n+\beta_n\ln X\bigr)\,,\qquad \alpha_4=\Omega\,,\quad\beta_5=-24\pi\,,\quad\beta_7=-\tfrac{158}{3}\,\pi\,,
\label{eq:boxlargeX}
\end{equation}
with $\beta_n=0$ for even $n$, where the expansion by regions gives the coefficients $\beta_n$ exactly. Exactly one solution in this space has no logarithms at large $X$. This solution is $v$ (normalized to begin with $X^{-4}$), and it turns out to be an elementary function,
\begin{equation}
v\;=\;\frac{4\sqrt{10}}{X\,\sqrt{(10X^2-1)\,(16X^4-10X^2+3)}}\,.
\label{eq:boxvclosed}
\end{equation}
The coefficients $\beta_n$, together with $\alpha_5$ and the three constants at the double poles in $\eps$ (appendix~\ref{app:uv}), then determine $u$, but they do not constrain the constant that multiplies $v$. This constant is fixed by one further condition. The function $v$ changes sign under continuation around the point $X^2=\tfrac1{10}$, where the loop integral is analytic, so the sum $u+\Omega\,v$ is single-valued there for only one value of the constant, which agrees with the integral in eq.~\eqref{eq:boxclosed} (appendix~\ref{app:uv}). The differential system therefore determines $\Omega$ as well. We have found no integer relation between $\Omega$ and known constants; appendix~\ref{app:largeX} writes it in terms of hypergeometric functions at rational arguments, eq.~\eqref{eq:Omegahyp}.

The expression $u+\Omega\,v$ thus contains no fitted numbers, and it agrees with the numerical values of $C_4$. It is, however, a closed form in a weaker sense than eq.~\eqref{eq:closedformgeneral}, since $u$ is an iterated integral over kernels that include the periods of three elliptic curves and of the K3 surface (appendix~\ref{app:largeX}) rather than a known function. For the coefficient $V_4$, the same construction leaves no undetermined constant (appendix~\ref{app:V4}).

\subsection{The box is not polylogarithmic}
\label{sec:boxmonodromy}

As the triangle shows, an elliptic curve among the master integrals does not by itself exclude a polylogarithmic coefficient. For the box, however, two independent arguments exclude it. The first concerns the function $D_0$, and it relies on the fact that iterated integrals of logarithmic forms with algebraic letters have quasi-unipotent monodromy, meaning that all eigenvalues of the monodromy matrices are roots of unity. We continue $D_0$ analytically in one of the lengths of the quadrilateral, around the points where the curve degenerates, and compute the continuation numerically. Its monodromy contains elements with a pair of eigenvalues $\lambda^{\pm1}$ for which $\lambda+\lambda^{-1}$ is an integer larger than two, so that $\lambda$ is not a root of unity (appendix~\ref{app:D0}). Hence $D_0$ is not an iterated integral of this type. As a control, the same computation for the triangle gives unipotent monodromy.

The second argument is exact and concerns the loop integrand. As the shape of the quadrilateral varies, the residue~\eqref{eq:boxresidue} and its derivatives obey a linear differential system. With one leg per site, the solutions of this system include the periods of the curve $Y^2=z_{23}\,z_{34}$, whereas for a logarithmic form on the same curve they do not. For the triangle, by contrast, the corresponding curve is a conic and has no periods.

Combined with eq.~\eqref{eq:softlimit}, the two arguments show that the periods of an elliptic curve enter the soft limit of the box with one leg per site. Furthermore, on the line of section~\ref{sec:boxline} the periods of three elliptic curves enter both $V_4$ and $C_4$. The box is therefore the first polygon of this theory that is not polylogarithmic.

\subsection{General momenta}
\label{sec:boxgeneral}

The results of sections~\ref{sec:boxline} and~\ref{sec:boxclosed} hold on a single line, for the regular tetrahedron. They are, however, restrictions of a structure in all variables of the box, namely the four site energies and the six lengths of the quadrilateral of momenta. Here we state its main properties, leaving a fuller treatment to a separate paper. The singular points of the master integrals lie on the zeros of 1333 irreducible polynomials in the site energies and the squares of the lengths, and every singular point in this section lies on one of them. The curves on which two partial energies vanish form two families of elliptic curves, one for pairs of partial energies that share an edge and one for pairs that share none. Their invariants are known in all variables, and on the line they reduce to the $j$-invariants of the three curves of section~\ref{sec:boxline}. The K3 surfaces of section~\ref{sec:boxk3} form a single family with six moduli, whose transcendental lattice has rank 5 for the regular tetrahedron, 6 for tetrahedra whose opposite edges have equal lengths, and 8 for generic momenta. In each case, the rank equals the order of the differential equation of the periods, so the operator~\eqref{eq:L5} and the form of the relation~\eqref{eq:quadrel} are special to the regular tetrahedron.

The elliptic structure at the partial-energy pole of a single site can also be made explicit for arbitrary momenta. In the coordinates $\sigma_1=y_{41}+y_{12}$, $\sigma_2=y_{12}+y_{23}$ and $\sigma_3=y_{23}+y_{34}$, which are the partial energies of sites 1, 2 and 3 up to the site energies, the measure $\dd^3\ell/(y_{12}y_{23}y_{34}y_{41})$ of the correlator becomes $\sqrt{g}\,\dd\sigma_1\,\dd\sigma_2\,\dd\sigma_3/\sqrt{\Delta}$, where $g$ is the Gram determinant of the quadrilateral and $\Delta$ is a polynomial of degree four in $\sigma_2$ whose coefficients are polynomials in $\sigma_1$, $\sigma_3$ and the six lengths (appendix~\ref{app:fibration}). The loop integral is therefore an integral over the plane of $\sigma_1$ and $\sigma_3$ of meromorphic differentials on the elliptic curve $w^2=\Delta(\sigma_2)$, with the site energies entering only through the rational weights, whose poles, the partial energies, are rational functions on the curve. The integration cycles and the two remaining integrations are not addressed here.

For one quadrilateral in general position, we have also reduced the correlator on the line $X_v=P_v+t$. The reduction contains an operator of order 8 in place of the fifth-order operator. The function $v$ of eq.~\eqref{eq:boxvclosed} is a rational function divided by the square root of a polynomial, and on this line the system has at most 16 solutions of this type. Writing $\Omega$ for the integral of eq.~\eqref{eq:boxclosed} with the edge energies of the given quadrilateral, we have not yet determined which combination of the 16 solutions multiplies $\Omega$, or whether other solutions multiply it as well. At large site energies, however, the leading term of $C_4$ is $\Omega/(X_1X_2X_3X_4)$ for all ratios of the site energies.

%% file: sec7_discussion.tex
\section{Conclusion}
\label{sec:discussion}

Bootstrapping loop corrections to cosmological correlators requires, first and foremost, knowing the class of functions in which they live. In flat space, decades of work on scattering amplitudes have mapped out this function space in considerable detail: one-loop amplitudes are polylogarithmic, and elliptic functions first appear at two loops. In cosmology the situation has been much less clear. The master integrals of the one-loop triangle involve an elliptic curve, which suggested that the triangle itself is elliptic, and that cosmological loops leave the space of polylogarithms one loop earlier than their flat-space counterparts.

In this paper, we have shown that this is not the case for the triangle. The elliptic curve appears in its master integrals, but the periods cancel in the physical combination, and the wavefunction coefficient of a conformally coupled scalar is a polylogarithmic function of transcendental weight two. We have given it in closed form both with one leg per site, eq.~\eqref{eq:closedform}, and for general site energies, eq.~\eqref{eq:closedformgeneral}, the latter being the input required for de~Sitter space and power-law cosmologies. Along the way, we identified the origin of the cancellation: at the poles of the elliptic sector, the residue of the loop integrand is a flat-space amplitude, rational in the loop momentum (proposition~\ref{prop:residues}). We also showed that the correlator integrand of the triangle has no residue involving an elliptic curve, though for a different reason, namely that it has no pole on the relevant planes at all.

The box, by contrast, is the first polygon of this theory that is not polylogarithmic, so cosmological loops do leave the space of polylogarithms at one loop, one graph later than expected. Its residue at the partial-energy pole of a single site involves two square roots rather than one. Its periods enter the wavefunction coefficient even with one leg per site, and in the correlator the elliptic periods do not cancel (section~\ref{sec:boxclosed}). For momenta forming a regular tetrahedron, the periods of a K3 surface enter the loop integrand as well (section~\ref{sec:box}). Taken together, the triangle and the box illustrate a broader lesson: the maximal cuts of the master integrals are not a reliable guide to the function space of the physical observable.

The triangle closed forms were bootstrapped rather than derived. The function space is fixed exactly on one line and by the symbol elsewhere, and the remaining rational coefficients were determined from numerical values of the integral. For one leg per site, eq.~\eqref{eq:closedform} can also be derived by carrying out the integrals over the edge energies analytically (appendix~\ref{app:derivation}). In this derivation the elliptic curve never appears, since the residue of proposition~\ref{prop:residues} reduces it to logarithms at the first integration, and with one leg per site the poles of the remaining integrals lie on lines along which every square root is rational. With several legs per site the poles move off these lines, and the radicands of eq.~\eqref{eq:closedformgeneral} are the values of the square roots at the points where they meet. For general site energies we have not carried the derivation through, but within the function space of section~\ref{sec:offline} the integrand fixes the ten coefficients of eq.~\eqref{eq:closedformgeneral} as residues (appendix~\ref{app:coefficients}). This confirms the bootstrap and explains why the coefficients are as simple as $-\tfrac14$, $-\tfrac12$, and $\pm1$. Relatedly, the curve of the master integrals lies on a factor that the correlator does not contain, and its periods also cancel in the wavefunction coefficient. These two facts trace back to the factorized term of the propagator~\eqref{eq:propagator}, but neither holds for the box, whose curves lie on poles of partial energies that the correlator also has. One natural place to look for a unifying principle is the cosmological polytope, in which the loop integrand is a canonical form while the master integrals are not. This formulation may help explain when the master integrals occupy a larger function space than the wavefunction coefficient.

It would also be valuable to carry these results into expanding backgrounds in full. With one leg per site, symmetry determines the de~Sitter triangle up to two numbers, the coefficient of the logarithm and a constant (section~\ref{sec:desitter}). With several legs at a site, however, the de~Sitter and power-law coefficients are integrals of eq.~\eqref{eq:closedformgeneral} over the site energies. Whether these integrals remain polylogarithmic (or whether the ten square roots can be rationalized simultaneously) is an open question.

For the box, the open problems concern the functions themselves. On one line, exact conditions on a differential system determine both the correlator and the coefficient with no fitted numbers. However, one of the two functions in the correlator is an iterated integral, of depth at most ten, over kernels that include the periods of three elliptic curves and of a K3 surface, rather than a known function. The number $\Omega$, a one-dimensional integral of elliptic integrals, can be written in terms of hypergeometric functions at rational arguments (appendix~\ref{app:largeX}), but has not been related to known constants. For generic momenta, the differential equation obeyed by the periods of the K3 surface has order 8, and while the reduction on a line contains this operator (section~\ref{sec:boxgeneral}), we have not constructed its solutions. Further afield, the residue of the pentagon at the partial-energy pole of a single site has three square roots and defines a curve of genus five. Beyond one loop, we have examined only integrands, and in the two-loop graphs of section~\ref{sec:census}, the correlator integrand contains partial energies only.

Finally, all fields in this paper are conformally coupled, which keeps the loop integrand rational. For fields of general mass, which are of particular interest for the cosmological collider program, the singular loci of the integrand are the same. However, the integrand is no longer rational, and our methods do not apply directly. The leading cosmological collider signals of a triangle loop of massive fields have recently been obtained using cutting rules \cite{Qin:2026yuo}, and one-loop signals of this kind can be integrated numerically \cite{Borinsky:2026roj}. The function space of such loops, however, remains open.

%% file: appA_method.tex
\section{Construction of the closed form}
\label{app:method}

In this appendix, we describe the construction of the closed form~\eqref{eq:closedform} in detail. We begin with the operator $\Lzero$ and its comparison with the integral, then construct its solutions, select the solution that equals $\Vc$ on the line~\eqref{eq:line}, and extend the result to a general triangle. Finally, we give three limits of the closed form, along with the four functions of eq.~\eqref{eq:closedformgeneral}.

\subsection{The operator and its singular points}
\label{app:operator}

The integration-by-parts identities of the family~\eqref{eq:family} follow from the vector fields that preserve $\Bk$ up to a factor. These are the polynomials $n_e$ and $n_0$ that solve the syzygy equation $\sum_en_e\,\partial\Bk/\partial y_e=n_0\,\Bk$ \cite{Benincasa:2024ptf}. For a rational function $g$ of the edge energies they give
\begin{equation}
\int\dd y_{12}\,\dd y_{23}\,\dd y_{31}\;\Bk^{-1/2+\eps}\,\Bigl[\,\sum_e\frac{\partial(n_e\,g)}{\partial y_e}+\bigl(\eps-\tfrac12\bigr)\,n_0\,g\Bigr]\;=\;0\,.
\label{eq:ibp}
\end{equation}
We solve these identities over finite fields at numerical values of $\lambda$ and $\eps$. Modulo the identities, the integrand of $\Vc$ and its derivatives with respect to $\lambda$ span a finite-dimensional space, and they therefore obey a linear relation: a differential operator of order 20 with coefficients that depend on $\lambda$ and $\eps$. Normalizing its leading coefficient to one, we reconstruct the remaining coefficients as rational functions of $\eps$ at fixed $\lambda$. All of them are finite at $\eps=0$, and their values there are the coefficients of $\Lzero$ up to this normalization; multiplying by the common denominator gives the polynomial coefficients $c_i(\lambda)$ of eq.~\eqref{eq:L0}. We then reconstruct these coefficients as rational functions of $\lambda$ and lift them to the rational numbers using the Chinese remainder theorem.

The operator $\Lzero$ of eq.~\eqref{eq:L0} has singular points at $\lambda=0$, $\pm\tfrac13$ and $\pm1$ and at infinity, and its local exponents are integers at each of them. The other zeros of its leading coefficient are apparent singularities, at which every solution is analytic. On the line~\eqref{eq:line} the finite singular points are the zeros of $\qG$, of the $\xi_i$ and of $X_2$,
\begin{equation}
\qG=3\lambda+1\,,\qquad \xi_1=1-\lambda\,,\qquad \xi_2=1+\lambda\,,\qquad \xi_3=3\lambda-1\,,\qquad X_2=\lambda\,.
\label{eq:lineletters}
\end{equation}
The points $\lambda=1$ and $\lambda=\tfrac13$ are the two folded configurations at the ends of the physical interval, and $\lambda=0$ is the limit in which two of the momenta become soft. The points $\lambda=-\tfrac13$ and $\lambda=-1$ lie outside the physical interval.

We note that if the limit $\eps\to0$ is taken before the reduction, the reduction has additional relations and gives an operator of order 14, which is a right factor of $\Lzero$ but does not annihilate $\Vc$. When the reduction is applied to a master integral of the elliptic sector, by contrast, its operator has a second-order factor equivalent to $\Ltwo$. A second reduction, which does not use the master integrals of section~\ref{sec:masters}, gives an operator of order 9 that divides $\Lzero$ from the right and annihilates $\Vc$. The structure of eq.~\eqref{eq:closedform} indicates that the operator of lowest order annihilating $\Vc(\lambda)$ has order 7, though we have not constructed it.

To test $\Lzero$ against numerical values of $\Vc$, we use a method that avoids differentiating them. We obtain the values from eq.~\eqref{eq:V} at $\eps=0$: the integral over $y_{12}$ is elementary, and we evaluate the remaining two-dimensional integral numerically to about 30 digits or better. Except for the 25 triangles of appendix~\ref{app:general} and the points of figure~\ref{fig:delta}, which use an independent evaluation to 11 digits, all comparisons with numerical values in section~\ref{sec:assembled} and in this appendix hold to this accuracy. Let $\varphi$ be a function on an interval $a\le\lambda\le b$ that vanishes at both ends together with its first 19 derivatives. Integration by parts then moves all derivatives from $\Vc$ to $\varphi$ without boundary terms,
\begin{equation}
\int_a^b\varphi\,\Lzero\Vc\;\dd\lambda \;=\; \int_a^b\Vc\,\Lzero^{\dagger}\varphi\;\dd\lambda\,,
\qquad
\Lzero^{\dagger}\varphi \;=\; \sum_{i=0}^{20}(-1)^i\,\frac{\dd^i}{\dd\lambda^i}\bigl(c_i\,\varphi\bigr)\,.
\label{eq:weakform}
\end{equation}
The right-hand side contains only the values of $\Vc$, and it vanishes for every such $\varphi$ if $\Lzero\Vc=0$. We take $\varphi_m=[(\lambda-a)(b-\lambda)]^{20}\,\mathrm{Leg}_m$, where $\mathrm{Leg}_m$ is the Legendre polynomial of degree $m$ on the interval. The right-hand side of eq.~\eqref{eq:weakform} vanishes to the accuracy of the numerical values for $\Lzero$, and it does not vanish for the operator of order 14.

\subsection{The space of solutions}
\label{app:solutions}

We now construct all solutions of $\Lzero$. Since the singular points are $0$, $\pm\tfrac13$ and $\pm1$, we look for solutions among the multiple polylogarithms with singular points at these five values. They are defined recursively by $G(;\lambda)=1$ and
\begin{equation}
\frac{\dd}{\dd\lambda}\,G(a,w;\lambda) \;=\; \frac{G(w;\lambda)}{\lambda-a}\,,
\label{eq:Gdef}
\end{equation}
where $a$ is a letter and $w$ is a word in the letters. The number of letters of a word is its weight. We make the ansatz
\begin{equation}
g(\lambda) \;=\; \sum_{w}R_w(\lambda)\,G(w;\lambda)\,,
\label{eq:mplansatz}
\end{equation}
where the $R_w$ are rational functions of $\lambda$ and the sum runs over all words up to a given weight. To apply $\Lzero$ to the ansatz we use the behavior of the multiple polylogarithms under a shift of the argument,
\begin{equation}
G(w;\lambda+\tau) \;=\; \sum_{w=w'w''}h_{w'}(\tau)\,G(w'';\lambda)\,,
\label{eq:pathcomposition}
\end{equation}
where the sum runs over all ways to split the word $w$ into a first part $w'$ and a second part $w''$. The function $h_{w'}(\tau)$ is the iterated integral of the word $w'$ along the segment from $\lambda$ to $\lambda+\tau$. It is a power series in $\tau$ whose coefficients are rational functions of $\lambda$. The $i$-th derivative of $g$ is $i!$ times the coefficient of $\tau^i$ in $g(\lambda+\tau)$, which we denote by $[\tau^i]$. We obtain
\begin{equation}
\Lzero\, g \;=\; \sum_{w''} G(w'';\lambda)\;\sum_{i=0}^{20}c_i\;i!\;\bigl[\tau^i\bigr]\sum_{w'}R_{w'w''}(\lambda+\tau)\,h_{w'}(\tau)\,.
\label{eq:recursion}
\end{equation}
Since the multiple polylogarithms are linearly independent over the rational functions, the coefficient of each $G(w'';\lambda)$ must vanish. The term with empty $w'$ is $\Lzero R_{w''}$, while the other terms contain only the coefficients of words longer than $w''$, so the coefficients of the longest words are rational solutions of $\Lzero$. The equations for the shorter words are then linear equations for rational functions, which we solve exactly, starting from the highest weight.

We find 5 rational solutions, 14 further solutions of weight one, and one solution of weight two. Since $\Lzero$ has order 20, these are all of its solutions, and the apparent singularities therefore do not contribute letters.

\subsection{The coefficient on the line}
\label{app:line}

The coefficient $\Vc$ is a combination of the 20 solutions with constant coefficients, which we determine in two steps. First, since the integral~\eqref{eq:V} is analytic at the folded configuration $\lambda=1$, we restrict to the solutions of $\Lzero$ that are analytic there, which form a subspace of dimension 14 that we determine exactly. Second, we fit the 14 constants of this subspace to numerical values of $\Vc$ at half of the points of our sample and identify them using an integer-relation algorithm \cite{Acres:2021sss}. Up to an overall factor of $\sqrt2\,\pi$, they are combinations of $1$, $\ln2$, $\ln^22$, and $\pi^2$ with rational coefficients. While the individual constants depend on the choice of basis in the subspace, the logarithms of 2 cancel in the final result, which agrees with the numerical values at the other half of the points.

All multiple polylogarithms of weight two in the result combine into the three dilogarithms $\mathrm{Li}_2(u_i)$ and into products of the logarithms $L_i=\ln(1-u_i)=\ln(2X_i/\qG)$. With the variables of eq.~\eqref{eq:lineletters} we obtain
\begin{equation}
\begin{aligned}
\frac{4\sqrt2\,\qG\,\xi_1^2\,\xi_2^2\,\xi_3^2}{\pi}\;\Vc
\;=\;{}& 2\lambda(1-\lambda)\,p_1\,\mathrm{Li}_2(u_1)+\lambda(1+\lambda)\,p_2\,\mathrm{Li}_2(u_2)\\
&+(3\lambda-1)\,p_3\,\mathrm{Li}_2(u_3)+\frac{4\pi^2}{3}\,\lambda^2\bigl(\lambda^2-6\lambda+1\bigr)\\
&+32\lambda^3\bigl[\,2\lambda\,L_1L_2+2\,L_1L_3+L_2L_3\bigr]\,,
\end{aligned}
\label{eq:lineformula}
\end{equation}
where the three polynomials are
\begin{equation}
p_1=9\lambda^3+25\lambda^2-\lambda-1\,,\qquad
p_2=-9\lambda^3+25\lambda^2+\lambda-1\,,\qquad
p_3=-3\lambda^3+15\lambda^2+3\lambda+1\,.
\label{eq:linepolys}
\end{equation}
The right-hand side of eq.~\eqref{eq:lineformula} vanishes to second order at $\lambda=1$ and at $\lambda=\tfrac13$, so $\Vc$ is finite at both ends of the physical interval. The ancillary files contain the 20 solutions in terms of multiple polylogarithms, the 14 constants in one basis, and the reduction of the result to eq.~\eqref{eq:lineformula}.

\subsection{General triangles}
\label{app:general}

For general momenta we have no differential operator, so we instead infer the result from the structure of eq.~\eqref{eq:lineformula}. In terms of the site energies $X=(2\lambda,\lambda,1)$, the prefactors of the dilogarithms are $X_i\,\xi_i$, and the three polynomials are given by a single expression,
\begin{equation}
p_i \;=\; \qG\,\xi_j\,\xi_k+8X_1X_2X_3\,,
\label{eq:pstructure}
\end{equation}
where $i$, $j$ and $k$ are distinct. For $i=1$ this expression is $(3\lambda+1)(1+\lambda)(3\lambda-1)+16\lambda^2$, which equals $p_1$. The coefficient of $L_i\,L_j$ is $16\,X_1X_2X_3\,X_iX_j$. This suggests the ansatz
\begin{equation}
\begin{aligned}
\frac{4\sqrt2\,\qG\,\xi_1^2\,\xi_2^2\,\xi_3^2}{\pi}\;\Vc
\;=\;{}& -\sum_{i}\xi_i\,b(X_i;X_j,X_k)\,\mathrm{Li}_2(u_i)+\pi^2\,c(X)\\
&+16\,X_1X_2X_3\sum_{i<j}X_iX_j\,L_i\,L_j
\end{aligned}
\label{eq:generalansatz}
\end{equation}
for an arbitrary triangle. Here $b(x;y,z)$ is a homogeneous polynomial of degree four that is symmetric in $y$ and $z$, and $c$ is a symmetric homogeneous polynomial of degree five. The polynomial $b$ has 9 coefficients and $c$ has 5. We restrict eq.~\eqref{eq:generalansatz} to the line and compare it with eq.~\eqref{eq:lineformula}. The three dilogarithms are linearly independent over the rational functions of $\lambda$, so the coefficient of each of them must agree as a polynomial in $\lambda$. This gives linear equations for the coefficients of $b$ and $c$. Their unique solution is
\begin{equation}
b(x;y,z) \;=\; -x\,\bigl[(x+y+z)(x-y+z)(x+y-z)+8\,x\,y\,z\bigr]\,,\qquad
c(X) \;=\; -\frac23\,X_1X_2X_3\sum_{i<j}\xi_i\,\xi_j\,.
\label{eq:bc}
\end{equation}
With these polynomials eq.~\eqref{eq:generalansatz} becomes eq.~\eqref{eq:closedform}.

While the comparison on the line determines the 14 coefficients, it does not test the form of the ansatz itself. We therefore compare eq.~\eqref{eq:closedform} with numerical values of the loop integral at 25 other triangles, which include two nearly folded and two soft configurations. With the coefficients fixed on the line, the formula agrees with the numerical values at all 25 triangles, to 11 digits, the precision of the evaluation used for this check. An independent fit of the 14 coefficients to numerical values returns the same polynomials, eq.~\eqref{eq:bc}. Equation~\eqref{eq:closedformgeneral} agrees with the numerical values on the line $X=P+\tfrac13(1,1,1)$ and at other points with $X_v\ge P_v$.

\subsection{Limits}
\label{app:limits}

We give three limits of eq.~\eqref{eq:closedform}. The first two are the smallest and the largest value of $\qG^2\,\Vc$ over all triangles of momenta. For an equilateral triangle with $X_v=X$ all three arguments are $u_i=\tfrac13$, and
\begin{equation}
\Vc \;=\; \frac{\sqrt2\,\pi}{8X^2}\Bigl[\,11\,\mathrm{Li}_2\bigl(\tfrac13\bigr)+16\ln^2\tfrac23-\frac{2\pi^2}{3}\Bigr]\,.
\label{eq:equilateral}
\end{equation}
When the momentum at site~1 becomes soft, the other two site energies approach a common value $X$. The result is finite and has no logarithm of $X_1$,
\begin{equation}
\Vc \;\longrightarrow\; \frac{3\sqrt2\,\pi}{128\,X^2}\,,\qquad X_1\to0\,.
\label{eq:softtriangle}
\end{equation}
For a folded configuration with $X_1=X_2+X_3$ the denominator of eq.~\eqref{eq:closedform} vanishes to second order in $\xi_1$. The numerator vanishes to the same order by the identity $\mathrm{Li}_2(x)+\mathrm{Li}_2(1-x)=\tfrac{\pi^2}{6}-\ln x\,\ln(1-x)$. With $X_{23}=X_2+X_3$ and $x=X_3/X_{23}$ the limit is
\begin{equation}
\Vc \;\longrightarrow\; \frac{\sqrt2\,\pi}{384\,X_{23}^2\,x^2(1-x)^2}\;N(x)\,,\qquad \xi_1\to0\,,
\label{eq:degenerate}
\end{equation}
\begin{equation}
\begin{aligned}
N(x)\;=\;{}&6\,(2x^3-3x^2+3x-1)\,\mathrm{Li}_2(x)+6\,x\,(x^2-x+1)\ln x\,\ln(1-x)-3\,x^2(x-2)\ln x\\
&+3\,(x^3-x^2-x+1)\ln(1-x)-\pi^2x^3-9\,x^2+9\,x\,.
\end{aligned}
\label{eq:degenerateN}
\end{equation}

\subsection{The functions of the general formula}
\label{app:functions}

Here we give the four functions of eq.~\eqref{eq:closedformgeneral}. For a square root $w$ and two polynomials $x$ and $y$ we write
\begin{equation}
\mathcal D[x;y] \;=\; \mathcal D\Bigl(\frac{x-w}{y-w}\,,\,\frac{x+w}{y+w}\Bigr)\,,
\label{eq:block}
\end{equation}
with $\mathcal D$ defined in eq.~\eqref{eq:BW}. The functions $F_0$ and $F_e$ are defined by tetrahedra. Vertices 1, 2 and 3 are the corners of the triangle of momenta, and vertex 4 is the loop momentum. For six lengths $\ell_{12},\ell_{13},\ell_{23},\ell_{14},\ell_{24},\ell_{34}$ the perimeters of the faces and of the closed paths are
\begin{equation}
\begin{aligned}
f_1&=\ell_{23}+\ell_{24}+\ell_{34}\,,&\quad f_2&=\ell_{13}+\ell_{14}+\ell_{34}\,,\\
f_3&=\ell_{12}+\ell_{14}+\ell_{24}\,,&\quad f_4&=\ell_{12}+\ell_{13}+\ell_{23}\,,\\
c_1&=\ell_{12}+\ell_{13}+\ell_{24}+\ell_{34}\,,&\quad c_2&=\ell_{12}+\ell_{14}+\ell_{23}+\ell_{34}\,,\\
c_3&=\ell_{13}+\ell_{14}+\ell_{23}+\ell_{24}\,.
\end{aligned}
\label{eq:perimeters}
\end{equation}
The polynomial $\mathcal P(h)$ of eq.~\eqref{eq:tetraquadratic} is quadratic, its discriminant is the radicand $R$, and $w=\sqrt R$. We write $p_a=-\mathcal P'(a)$, so that $\mathcal D[p_a;p_0]$ is the function $\mathcal D_a$ of section~\ref{sec:offline}. For $F_0$ the lengths are
\begin{equation}
(\ell_{12},\ell_{13},\ell_{23})=(P_2,P_1,P_3)\,,\qquad (\ell_{14},\ell_{24},\ell_{34})=\tfrac12\bigl(X_1+X_2-X_3,\;X_2+X_3-X_1,\;X_1+X_3-X_2\bigr)\,,
\label{eq:lengthsF0}
\end{equation}
and $F_0$ is given in eq.~\eqref{eq:F0}. For the edge $e=12$ the lengths are
\begin{equation}
(\ell_{12},\ell_{13},\ell_{23})=(P_2,P_1,P_3)\,,\qquad (\ell_{14},\ell_{24},\ell_{34})=\bigl(\tfrac12\qG,\;\tfrac12\qG-X_2,\;\tfrac12\qG-X_1\bigr)\,.
\label{eq:lengthsFe}
\end{equation}
A prime denotes the same quantities with $\ell_{24}\to-\ell_{24}$ and $\ell_{34}\to-\ell_{34}$, which has the same radicand. Then
\begin{equation}
\begin{aligned}
F_{12} \;=\;{}& -\mathcal D[p_{f_1};p_0]+\mathcal D[p_{c_1};p_0]+\sum_{k=2}^{4}\mathcal D[p'_{f_k};p'_0]-\mathcal D[p'_{c_2};p'_0]-\mathcal D[p'_{c_3};p'_0]\\
&+\mathcal D[p'_0;p_{f_4}]-\mathcal D[p'_{f_4};p_0]\,.
\end{aligned}
\label{eq:Fe}
\end{equation}
The functions $F^\pm_{12}$ have $w=\sqrt{T^\pm_{12}}$,
\begin{equation}
\begin{aligned}
F^-_{12} \;&=\; \mathcal D[a_1;a_2]+\mathcal D[a_3;a_7]+\mathcal D[a_4;a_7]-\mathcal D[a_5;a_6]\,,\\
F^+_{12} \;&=\; -\mathcal D[b_1;b_2]+\mathcal D[b_1;b_5]+\mathcal D[b_1;b_6]+\mathcal D[b_3;b_4]-\mathcal D[b_3;b_5]\,,
\end{aligned}
\label{eq:Fpm}
\end{equation}
with the polynomials
\begin{equation}
\begin{aligned}
a_1&=P_2^2-(P_1+P_3)^2+2X_1(P_1+P_3)-2P_1X_2\,,\\
a_2&=P_2^2-P_1^2-P_3^2+2X_1(X_1-X_2)\,,\qquad a_3=P_3^2-(P_1-P_2)^2+2P_1X_2-2P_2X_1\,,\\
a_4&=P_2^2+P_3^2-P_1^2+2X_2(X_1-X_2)\,,\qquad a_6=P_3^2-(P_1-P_2)^2-2P_1X_2+2P_2X_1\,,\\
a_5&=(P_2+P_3)^2-P_1^2+2X_1P_2-2X_2(P_2+P_3)\,,\\
a_7&=P_2^2-(P_1+P_3)^2-2X_1(P_1+P_3)+2P_1X_2\,,\\
b_1&=P_3^2-P_1^2-P_2^2-2X_1X_2\,,\qquad b_2=P_3^2-(P_1-P_2)^2-2P_1X_2-2P_2X_1\,,\\
b_3&=P_2^2-(P_1+P_3)^2+2X_1(P_1+P_3)+2P_1X_2\,,\\
b_4&=P_2^2-P_1^2-P_3^2+2X_1(X_1+X_2)\,,\qquad b_5=P_2^2+P_3^2-P_1^2-2X_2(X_1+X_2)\,,\\
b_6&=(P_2+P_3)^2-P_1^2+2X_1P_2+2X_2(P_2+P_3)\,.
\end{aligned}
\label{eq:abpolys}
\end{equation}
The other functions follow from relabelings of the sites. The functions $F_{23}$ and $F^\pm_{23}$ follow from $F_{12}$ and $F^\pm_{12}$ under $(1,2,3)\to(2,3,1)$, applied to the $X_v$ and to the $P_v$. The functions $F_{31}$ and $F^\pm_{13}$ follow under $(1,2,3)\to(1,3,2)$.

%% file: appB_box.tex
\section{Derivations for the box}
\label{app:box}

In this appendix, we derive the results of section~\ref{sec:box}. We first derive the residue at the partial-energy pole of a single site and the geometry of the threshold, and then reduce $D_0$ to Carlson integrals. We then give the periods of the elliptic curves on the line of section~\ref{sec:boxline}, an elliptic fibration of the loop measure for general momenta, the fifth-order operator, the conditions that determine $u$ and $v$, and the expansions of $C_4$ and $V_4$ at large site energy.

\subsection{The residue at the partial-energy pole of a single site}
\label{app:boxresidue}

The pole at $X_1+y_{41}+y_{12}=0$ comes from the region of the time integrals in which $t_1$ is larger than the times of the two neighboring sites. For $t_1>t_2$ the propagator~\eqref{eq:propagator} factorizes,
\begin{equation}
G_{y_{12}}(t_1,t_2) \;=\; \frac{e^{-y_{12}t_1}}{2y_{12}}\,\bigl(e^{\,y_{12}t_2}-e^{-y_{12}t_2}\bigr)\,,
\label{eq:propfactor}
\end{equation}
and the same holds for $G_{y_{41}}(t_1,t_4)$. The integral over $t_1$ from the larger of $t_2$ and $t_4$ to infinity is then
\begin{equation}
\int_{\max(t_2,t_4)}^{\infty}\!\dd t_1\;e^{-(X_1+y_{41}+y_{12})\,t_1} \;=\; \frac{e^{-(X_1+y_{41}+y_{12})\max(t_2,t_4)}}{X_1+y_{41}+y_{12}}\,.
\label{eq:t1integral}
\end{equation}
At the pole the exponential on the right-hand side equals one, and the residue is an integral over the times of the other three sites. Each of the two brackets of the form~\eqref{eq:propfactor} adds $\pm y$ to the energy of a neighboring site, with the sign as a factor. The remaining integral is that of the chain of sites 2, 3 and 4, which gives eq.~\eqref{eq:boxresidue}. The recursion for $\psi_4$ gives the same residue. For the triangle the same steps give eq.~\eqref{eq:triangleresidue}.

\subsection{The threshold}
\label{app:threshold}

At the pole the sum of the distances from $\vec\ell$ to two fixed points is constant. The points are those at which $y_{12}$ and $y_{41}$ vanish, and they are a distance $P_1$ apart. The loop momentum therefore lies on a spheroid with these points as foci. At the threshold $y_{41}+y_{12}=P_1$ the spheroid becomes the segment between the foci, on which $y_{12}=\tfrac12(P_1-\chi)$ and $y_{41}=\tfrac12(P_1+\chi)$ with $\chi=y_{41}-y_{12}$. The distance from a point of the segment to a third point follows from Stewart's theorem. Let $C$ be a point at distance $a$ from the focus at which $y_{12}$ vanishes and at distance $b$ from the other focus. The theorem gives
\begin{equation}
|\vec\ell-\vec C|^2 \;=\; \frac{(P_1+\chi)\,a^2+(P_1-\chi)\,b^2}{2P_1}-\frac{P_1^2-\chi^2}{4}\,.
\label{eq:stewart}
\end{equation}
With the two remaining vertices of the quadrilateral in place of $C$ this is eq.~\eqref{eq:segment}. Both $z_{23}=y_{23}^2$ and $z_{34}=y_{34}^2$ are quadratic in $\chi$ with leading coefficient $\tfrac14$. We write
\begin{equation}
z_{23} \;=\; \tfrac14\,(\chi-e_1)(\chi-e_2)\,,\qquad z_{34} \;=\; \tfrac14\,(\chi-e_3)(\chi-e_4)\,,\qquad 16\,Y^2 \;=\; \prod_{i=1}^{4}(\chi-e_i)\,,
\label{eq:roots}
\end{equation}
so that $Y^2=z_{23}\,z_{34}$. The roots are
\begin{equation}
e_{1,2} \;=\; -m\pm\sqrt{m^2-n}\,,\qquad m=\frac{P_2^2-P_{12}^2}{P_1}\,,\qquad n=2\bigl(P_2^2+P_{12}^2\bigr)-P_1^2\,,
\label{eq:rootvalues}
\end{equation}
and $e_{3,4}$ follow with $P_2\to P_{41}$ and $P_{12}\to P_4$. For generic momenta the roots form two distinct complex-conjugate pairs; they coincide for symmetric configurations such as the regular tetrahedron.

\subsection{Reduction of \texorpdfstring{$D_0$}{D\_0} to Carlson integrals}
\label{app:D0}

The integrand of eq.~\eqref{eq:D0} is a rational function of $\chi$, $y_{23}$ and $y_{34}$. We separate it according to its parity under $y_{23}\to-y_{23}$ and $y_{34}\to-y_{34}$. The part that is even in both square roots is rational in $\chi$. The parts that are odd in one square root contain a single square root of a quadratic polynomial, and they integrate to logarithms. The part that is odd in both square roots is a rational function of $\chi$ divided by $Y$, and it gives the elliptic integrals.

The chain coefficient~\eqref{eq:chain} has poles where one of six linear factors vanishes. Each zero lies at a point $\chi=\alpha$ on one of the four sheets of the two square roots. We label the sheets by the values $\eta=\pm y_{23}(\alpha)$ and $\eta'=\pm y_{34}(\alpha)$ of the roots at the pole. For momenta that do not lie in a plane all these poles are simple. Let $\rho$ be the residue of the integrand at such a pole. The partial-fraction decomposition then gives eq.~\eqref{eq:D0closed},
\begin{equation}
D_0 \;=\; \frac{\pi}{4P_1}\biggl[\;\frac14\sum_{(\alpha,\eta,\eta')}\rho\,\Bigl(L_0+\eta\,L_{23}+\eta'L_{34}+\eta\,\eta'\,L_\Pi\Bigr)+4\,p_0\,J_F\biggr]\,,
\label{eq:D0full}
\end{equation}
with the integrals
\begin{equation}
L_0=\int_{-P_1}^{P_1}\frac{\dd\chi}{\chi-\alpha}\,,\qquad
L_{23}=\int_{-P_1}^{P_1}\frac{\dd\chi}{(\chi-\alpha)\,y_{23}}\,,\qquad
L_\Pi=\int_{-P_1}^{P_1}\frac{\dd\chi}{(\chi-\alpha)\,y_{23}\,y_{34}}\,,
\label{eq:fourintegrals}
\end{equation}
and $L_{34}$ is defined as $L_{23}$ with $y_{34}$ in place of $y_{23}$. The part that is odd in both square roots also contains a constant divided by $Y$, whose integral is proportional to the integral of the first kind $J_F=\int\dd\chi/(4Y)$. The coefficient of $J_F$ follows from the partial fractions at the branch point $\chi=e_1$,
\begin{equation}
p_0 \;=\; \frac14\sum_{(\alpha,\eta,\eta')}\frac{\rho\,\eta\,\eta'}{\alpha-e_1}\,.
\label{eq:p0}
\end{equation}

The first two integrals in eq.~\eqref{eq:fourintegrals} are elementary,
\begin{equation}
\begin{aligned}
L_0&=\ln\frac{P_1-\alpha}{-P_1-\alpha}\,,\\
L_{23}&=-\frac{1}{\sqrt{z_{23}(\alpha)}}\,\biggl[\ln\frac{2\,z_{23}(\alpha)+z_{23}'(\alpha)\,(\chi-\alpha)+2\sqrt{z_{23}(\alpha)\,z_{23}(\chi)}}{\chi-\alpha}\biggr]_{\chi=-P_1}^{\chi=P_1}\,.
\end{aligned}
\label{eq:elementary}
\end{equation}

For the elliptic integrals we use the symmetric integrals of Carlson \cite{Carlson:1995}, which are $R_F(a,b,c)=\tfrac12\int_0^\infty\dd\tau\,[(\tau+a)(\tau+b)(\tau+c)]^{-1/2}$ and $R_J(a,b,c,p)=\tfrac32\int_0^\infty\dd\tau\,(\tau+p)^{-1}[(\tau+a)(\tau+b)(\tau+c)]^{-1/2}$. With $r_i=\sqrt{P_1-e_i}$ and $\bar r_i=\sqrt{-P_1-e_i}$ we define
\begin{equation}
U_{ij} \;=\; \frac{r_i\,r_j\,\bar r_k\,\bar r_l+\bar r_i\,\bar r_j\,r_k\,r_l}{2P_1}\,,
\label{eq:Uij}
\end{equation}
where $i$, $j$, $k$ and $l$ are distinct. The integral of the first kind is
\begin{equation}
J_F \;=\; \int_{-P_1}^{P_1}\frac{\dd\chi}{4\,Y} \;=\; 2\,R_F\bigl(U_{12}^2,U_{13}^2,U_{14}^2\bigr)\,.
\label{eq:JF}
\end{equation}
Since $Y=y_{23}\,y_{34}$, the integral of the third kind is $L_\Pi=4\,(J_\Pi-J_F)/(\alpha-e_1)$ with
\begin{equation}
J_\Pi \;=\; \int_{-P_1}^{P_1}\frac{(\chi-e_1)\,\dd\chi}{4\,(\chi-\alpha)\,Y}
\;=\; \frac{2\,d_{12}\,d_{13}\,d_{14}}{3\,d_{15}}\;R_J\bigl(U_{12}^2,U_{13}^2,U_{14}^2,U_{15}^2\bigr)+\frac1w\,\ln\frac{S_5+w}{S_5-w}\,,
\label{eq:JPi}
\end{equation}
where $d_{ij}=e_j-e_i$ with $e_5=\alpha$, and
\begin{equation}
\begin{aligned}
U_{15}^2 &= U_{12}^2-\frac{d_{13}\,d_{14}\,d_{25}}{d_{15}}\,,\qquad
w^2 = S_5^2+\frac{(P_1^2-\alpha^2)\,U_{15}^2}{r_1^2\,\bar r_1^2}\,,\\
S_5 &= \frac{1}{2P_1}\biggl[\frac{r_2\,r_3\,r_4}{r_1}\,(-P_1-\alpha)+\frac{\bar r_2\,\bar r_3\,\bar r_4}{\bar r_1}\,(P_1-\alpha)\biggr]\,.
\end{aligned}
\label{eq:S5}
\end{equation}

Equations~\eqref{eq:D0full} to~\eqref{eq:S5} are the complete result. We determine the branches of the logarithms and of the Carlson integrals by continuation from $\chi=-P_1$. The formulas in terms of $R_F$ and $R_J$ hold on a part of the segment on which the real quantities $U_{ij}$ are positive. If one of them changes sign on the segment, we divide the segment at that point and add the integrals of the parts. Poles at the end points $\alpha=\pm P_1$ contribute their finite parts, and real poles inside the segment contribute principal values. The ancillary files contain a numerical implementation of these rules. 

For momenta in a plane the factor $P_3+y_{23}+y_{34}$ has a double zero at a point $\chi=\alpha_*$ on one sheet. On this sheet the integrand has the expansion $a_*/(\chi-\alpha_*)^2+b_*/(\chi-\alpha_*)$ up to regular terms. We write $F(\alpha)=L_0+\eta\,L_{23}+\eta'L_{34}+\eta\,\eta'L_\Pi$ for the bracket of eq.~\eqref{eq:D0full}, where $\eta$ and $\eta'$ are the values of the square roots at $\chi=\alpha$ on the sheet. Since $\int\dd\chi\,g(\chi)/(\chi-\alpha)^2$ is the derivative of $\int\dd\chi\,g(\chi)/(\chi-\alpha)$ with respect to $\alpha$, the double pole contributes
\begin{equation}
\rho\,F(\alpha)\;\longrightarrow\; a_*\,\frac{\dd F}{\dd\alpha}(\alpha_*)+b_*\,F(\alpha_*)\,,\qquad
\frac{\rho\,\eta\,\eta'}{\alpha-e_1}\;\longrightarrow\; a_*\,\frac{\dd}{\dd\alpha}\biggl[\frac{\eta\,\eta'}{\alpha-e_1}\biggr]_{\alpha_*}+b_*\,\frac{\eta\,\eta'}{\alpha_*-e_1}
\label{eq:doublepole}
\end{equation}
in eqs.~\eqref{eq:D0full} and~\eqref{eq:p0}. The derivatives are taken along the sheet. The ancillary implementation includes this term.

We obtain the monodromy of $D_0$ of section~\ref{sec:boxmonodromy} by numerical continuation in one of the lengths of the quadrilateral. We restrict it to the minimal local system of $D_0$, the span of its analytic continuations, which has dimension 5 for each of two quadrilaterals that do not lie in a plane (9 with several legs per site). In both cases it contains elements with a single pair of eigenvalues $\lambda^{\pm1}$ off the unit circle, with $\lambda+\lambda^{-1}=4$, 6, 18 and 20 to 30 digits; for the element with $\lambda+\lambda^{-1}=18$, for instance, $\lambda=9+4\sqrt5$. The transport matrices along the loops, the loop paths and the script that extracts the eigenvalues are in the ancillary files. Since $|\lambda|>1$, these eigenvalues are not roots of unity, whatever the dimension of the representation. The differential system of the residue~\eqref{eq:boxresidue} has rank 6 on one family of quadrilaterals and rank 4 on a second family. In both cases its solutions include the two periods of the curve. The system of a logarithmic form on the same curve has rank 1.

\subsection{The elliptic curves on the line}
\label{app:curves}

Three elliptic curves appear on the line of section~\ref{sec:boxline}, namely the curves on which two partial energies vanish. The two partial energies may be those of two adjacent sites, of two opposite sites, or of one site and the pair of adjacent sites that contains it. We denote the curves of these three cases by $E_{\rm adj}$, $E_{\rm opp}$ and $E_{\rm nest}$, in this order.

The periods of all three are hypergeometric functions of the $j$-invariant. We write a curve in the Weierstrass form $Y_E^2=4X_E^3-g_2\,X_E-g_3$. Under the rescaling $g_2\to\nu^4g_2$ and $g_3\to\nu^6g_3$ its periods are multiplied by $\nu^{-1}$. A period is therefore $g_2^{-1/4}$ times a function of the invariant ratio
\begin{equation}
J \;=\; \frac{1728}{j} \;=\; 1-\frac{27\,g_3^2}{g_2^3}\,.
\label{eq:Jdef}
\end{equation}
This function obeys the hypergeometric equation with parameters $\tfrac1{12}$, $\tfrac5{12}$ and $1$. For $0<J<1$ and $g_3>0$ the real period is
\begin{equation}
\varpi_E \;=\; \frac{2\pi}{(12\,g_2)^{1/4}}\;{}_2F_1\bigl(\tfrac1{12},\tfrac5{12};1;J\bigr)\,.
\label{eq:period}
\end{equation}
For $g_3<0$ the same expression is the absolute value of the imaginary period. In both cases it is the period of smallest absolute value.

On the line $j$ is a rational function of degree 24 in $t$ for each curve. We denote by $c_4$ the polynomial of degree eight whose cube is the numerator of $j$. In a Weierstrass form with polynomial coefficients, $g_2$ is proportional to $c_4$. The Picard--Fuchs operator $\mathcal L_E$ of each curve is therefore the hypergeometric operator in the variable $J(t)$, conjugated by $c_4^{1/4}$,
\begin{equation}
\mathcal L_E \;=\; c_4^{-1/4}\,\Bigl[\,J(1-J)\,\frac{\dd^2}{\dd J^2}+\Bigl(1-\frac{3J}{2}\Bigr)\frac{\dd}{\dd J}-\frac{5}{144}\,\Bigr]\,c_4^{1/4}\,,
\label{eq:boxhyp}
\end{equation}
with both sides normalized to unit leading coefficient in $\dd/\dd t$. Equation~\eqref{eq:boxhyp} holds exactly for the three operators of the reduction. 
The point $J=0$ is $j=\infty$, where a curve degenerates. This happens at $t=0$ for all three curves and at $t=-1+\sqrt{3/2}$ for $E_{\rm opp}$. The variable $t$ is not a modular function for any of the three curves. Each map $t\mapsto j$ has critical points away from $j=0$, $1728$ and $\infty$. These critical points are the apparent singularities of $\mathcal L_E$. We give the three $j$-invariants in the ancillary files.

\subsection{An elliptic fibration for general momenta}
\label{app:fibration}

For arbitrary momenta, the measure of the correlator can be written as an elliptic fibration in coordinates built from the edge energies. Let $\vec A_e$ be the vertices of the quadrilateral of momenta, so that $y_e=|\vec\ell-\vec A_e|$, and put $\vec e_1=\vec A_{41}-\vec A_{12}$, $\vec e_2=\vec A_{23}-\vec A_{12}$, $\vec e_3=\vec A_{34}-\vec A_{12}$, with Gram matrix $G_{ij}=\vec e_i\cdot\vec e_j$ and $g=\det G$. The diagonal entries are $P_1^2$, $P_2^2$ and the square of the length of the diagonal $\vec A_{12}\vec A_{34}$, and $g$ vanishes only for momenta in a plane. Define
\begin{equation}
\sigma_1=y_{41}+y_{12}\,,\qquad \sigma_2=y_{12}+y_{23}\,,\qquad \sigma_3=y_{23}+y_{34}\,,
\label{eq:uvs}
\end{equation}
and the vectors $L=(\sigma_1,\,\sigma_2,\,\sigma_2-\sigma_3)$ and $m=\tfrac12\bigl(G_{11}-\sigma_1^2,\;G_{22}-\sigma_2^2,\;G_{33}-(\sigma_2-\sigma_3)^2\bigr)$. The components $\eta_i=\vec\ell\cdot\vec e_i$ of the loop momentum, measured from $\vec A_{12}$, are $\eta=L\,y_{12}+m$, since for instance $y_{41}^2=y_{12}^2-2\eta_1+G_{11}$ and $y_{41}=\sigma_1-y_{12}$. The condition $|\vec\ell\,|^2=y_{12}^2$, that is $\eta^{\rm T}G^{-1}\eta=y_{12}^2$, then becomes a quadratic equation for $y_{12}$,
\begin{equation}
\mathcal A\,y_{12}^2+2\,\mathcal B\,y_{12}+\mathcal C=0\,,\qquad
\mathcal A=L^{\rm T}\operatorname{adj}(G)\,L-g\,,\quad
\mathcal B=L^{\rm T}\operatorname{adj}(G)\,m\,,\quad
\mathcal C=m^{\rm T}\operatorname{adj}(G)\,m\,,
\label{eq:fibquadratic}
\end{equation}
so that $y_{12}=(-\mathcal B\pm\sqrt{\Delta})/\mathcal A$ with
\begin{equation}
\Delta(\sigma_2;\sigma_1,\sigma_3)\;=\;\mathcal B^2-\mathcal A\,\mathcal C\,.
\label{eq:fibdelta}
\end{equation}
Although $\mathcal A$, $\mathcal B$ and $\mathcal C$ have degree up to four in $\sigma_2$, the combination $\Delta$ has degree exactly four in $\sigma_2$, and also in $\sigma_1$ and in $\sigma_3$, for independent entries of $G$. For fixed $\sigma_1$ and $\sigma_3$ the equation $w^2=\Delta(\sigma_2)$ therefore defines an elliptic curve, and the other three edge energies $y_{23}=\sigma_2-y_{12}$, $y_{34}=\sigma_3-\sigma_2+y_{12}$ and $y_{41}=\sigma_1-y_{12}$ are rational functions of $\sigma_2$ and $w$ on it. Computing the Jacobian of the map $\vec\ell\mapsto(\sigma_1,\sigma_2,\sigma_3)$ gives, on each branch of the square root on which the four edge energies are positive,
\begin{equation}
\frac{\dd^3\ell}{y_{12}\,y_{23}\,y_{34}\,y_{41}}\;=\;\frac{\sqrt{g}}{\sqrt{\Delta}}\;\dd\sigma_1\,\dd\sigma_2\,\dd\sigma_3\,.
\label{eq:fibmeasure}
\end{equation}
Both statements hold as identities for independent Gram entries; we have checked them symbolically, and eq.~\eqref{eq:fibmeasure} numerically at generic points. The correlator integrand is this measure times a rational function of the site energies and the edge energies, with poles at the partial energies: $X_1+\sigma_1$, $X_2+\sigma_2$, $X_3+\sigma_3$ and $X_4+\sigma_1-\sigma_2+\sigma_3$ for the single sites, and, for pairs and triples of adjacent sites, linear functions of $y_{12}$, which are rational functions of $\sigma_2$ and $w$ on the curve. The loop integral is thus a two-dimensional integral over $\sigma_1$ and $\sigma_3$ of meromorphic differentials on the curves $w^2=\Delta(\sigma_2;\sigma_1,\sigma_3)$, and in particular the number $\Omega$ of eq.~\eqref{eq:boxclosed} is an integral of elliptic integrals over the plane of $\sigma_1$ and $\sigma_3$. The positivity constraints on the edge energies select the intervals of $\sigma_2$ and the branches, and the integration cycles for the energy-dependent weights, as well as the two remaining integrations, have not been worked out.

\subsection{The fifth-order operator}
\label{app:L5}

The order of the differential equation of the periods is 5 for the regular tetrahedron, 6 for tetrahedra whose opposite edges have equal lengths, and 8 for generic momenta. The fifth-order operator of section~\ref{sec:boxk3} arises in three ways. It is the operator of the periods of the surface as a function of $\Sigma=y_{41}+y_{12}$. It is the operator of the same surfaces along the line $X_v=1+t$, on which the pole of site~1 lies at $\Sigma=-(1+t)$. It is also the irreducible factor of order 5 of the reduction on that line. With the substitution $\Sigma\to1+t$ the three operators are identical. The solutions have no logarithms at infinite $\Sigma$.

The operator is $\mathcal L_5=\sum_{i=0}^{5}q_i(\Sigma)\,\dd^i/\dd\Sigma^i$ with
\begin{equation}
\begin{aligned}
q_5&=\Sigma^2(\Sigma^2-1)^2(3\Sigma^2+1)(\Sigma^4-4\Sigma^2+12)(26\Sigma^4-41\Sigma^2+47)\,,\\
q_4&=\Sigma\,(\Sigma^2-1)\bigl(2184\Sigma^{12}-10788\Sigma^{10}+30503\Sigma^8-38330\Sigma^6+33139\Sigma^4-2272\Sigma^2+1692\bigr)\,,\\
q_3&=19578\Sigma^{14}-93867\Sigma^{12}+232539\Sigma^{10}-318461\Sigma^8+274863\Sigma^6-70908\Sigma^4-3964\Sigma^2+1692\,,\\
q_2&=3\,\Sigma^3\bigl(21918\Sigma^{10}-83741\Sigma^8+168795\Sigma^6-189111\Sigma^4+115763\Sigma^2-7896\bigr)\,,\\
q_1&=18\,\Sigma^2\bigl(4082\Sigma^{10}-12389\Sigma^8+22570\Sigma^6-19379\Sigma^4+5392\Sigma^2+1316\bigr)\,,\\
q_0&=90\,\Sigma^5\bigl(182\Sigma^6-447\Sigma^4+846\Sigma^2-509\bigr)\,.
\end{aligned}
\label{eq:L5}
\end{equation}
In a suitable basis the five solutions $\Pi_1,\ldots,\Pi_5$ obey
\begin{equation}
\Pi_1\,\Pi_2+\Pi_3\,\Pi_4+\Pi_5^2\;=\;0\,.
\label{eq:quadrel}
\end{equation}
The symmetric square of $\mathcal L_5$ therefore has order 14 instead of 15. A relation of this form holds for the periods of a K3 surface whose transcendental lattice has rank 5.

The solutions of the fifth-order operator enter the discontinuity of $V_4$. For momenta that form a regular tetrahedron of unit side these solutions have branch points at the roots $\Sigma_c$ of $\Sigma^4-4\Sigma^2+12$, where they behave as $(\Sigma-\Sigma_c)^{1/2}$. The residue~\eqref{eq:boxresidue} vanishes to second order at the point where two singular points of the surface coincide, so a function that contains the solutions through this residue has a term proportional to $(\Sigma-\Sigma_c)^{5/2}$, with a coefficient fixed by the residue there. With one leg per site the discontinuity of $V_4$ has this branch point, with a numerical exponent of $2.507$ and with the predicted coefficient. With several legs per site the leading term of the residue at large site energies generates all five solutions, whereas a form of the opposite parity generates none of them. With one leg per site the differential system of the loop integrand of $V_4$ has rank 36, and that of the correlator has rank 34. These systems are distinct from the quotient systems of dimension 34 and 28 on the line of appendix~\ref{app:uv}. The quotient of the former by the other forms at the partial-energy pole of a single site has rank 5, and its operator is the fifth-order operator.

\subsection{The solutions on the line}
\label{app:uv}

The reduction on the line gives first-order systems for the master integrals and their coefficients in the expansion in $\eps$. Since the master integrals have double poles in $\eps$, the systems that include the double poles have dimension 81 for $V_4$ and 67 for $C_4$. The coefficient of the double pole of each master integral is a residue at infinite loop momentum, equal to $0$, $-\pi/8$, or $-\pi/16$, and with these values $V_4$ and $C_4$ are solutions of quotient systems of dimension 34 and 28. Each second-order operator of an elliptic curve occurs twice, once for each order of the pole in $\eps$, but only the one at the simple pole contributes to $V_4$ and $C_4$.

We compare the systems with numerical values of both functions. Some care is needed here, since a fit of all boundary constants to values on a short interval would reproduce the values of any smooth function, and so would not test the systems. We therefore use values close to the singular points and at large $t$, together with the conditions on the double poles. The solutions then reproduce the values that were not used in the fit, while the same fit to a different function fails.

An exact rational change of basis brings the system of dimension 28 to a form with simple poles only. The poles lie at the zeros of the 19 polynomials
\begin{equation}
\begin{aligned}
&3X^{2}+1,\quad 10X^{2}-1,\quad X^{4}-4X^{2}+12,\quad 16X^{4}-10X^{2}+3,\\
&X,\quad X+2,\quad X-2,\quad 3X+1,\quad 3X-1,\quad 2X^{2}+1,\quad 13X^{2}-4,\quad 12X^{4}-20X^{2}+9,\\
&X+1,\quad X-1,\quad 2X+1,\quad 2X-1,\quad 2X^{2}-1,\quad 2X^{2}-3,\quad 4X^{4}-4X^{2}+3\,,
\end{aligned}
\label{eq:letters}
\end{equation}
and the system has no other singular points at finite $X$. At the zeros of the polynomials in the first line the solutions have square-root branch points, and at those of the second line they have logarithms. At those of the third line both occur.

The loop integral converges for $t>-1$, so $C_4$ is single-valued around the singular points of the system in that range. In the system of dimension 28 the solutions that are single-valued around the six points
\begin{equation}
t \;=\; 1\,,\quad -1+\sqrt{\tfrac32}\,,\quad 0\,,\quad -1+\tfrac{1}{\sqrt2}\,,\quad -1+\tfrac{2}{\sqrt{13}}\,,\quad -\tfrac12
\label{eq:sixpoints}
\end{equation}
form a space of dimension 4. Those that are also single-valued around the logarithmic point $t=-\tfrac23$ form a space of dimension 3. For $V_4$ the corresponding dimensions are 8 and 7. The remaining singular point in this range is $t=-1+1/\sqrt{10}$. Exactly one solution in these spaces has no logarithms at large $X$. This solution is the function $v$, which we normalize to begin with $X^{-4}$. It is the elementary function of eq.~\eqref{eq:boxvclosed}, and its expansion is even in $1/X$ with rational coefficients,
\begin{equation}
v\;=\;X^{-4}+\frac{29}{80}\,X^{-6}+\frac{923}{12800}\,X^{-8}-\frac{1531}{204800}\,X^{-10}+O\bigl(X^{-12}\bigr)\,.
\label{eq:boxv}
\end{equation}
The function $u$ has $\alpha_4=0$ and the coefficients $\beta_4$ to $\beta_8$ and $\alpha_5$ of appendix~\ref{app:largeX}. Its three boundary constants at the double poles in $\eps$ equal $-\pi/8$. None of these conditions constrains the constant that multiplies $v$, and the condition at $t=-\tfrac23$ does not constrain it either.

The condition at $t=-1+1/\sqrt{10}$ determines this constant. At this point $X^2=\tfrac1{10}$, and $v$ has a square-root branch point. Continuation around the point gives $v\to-v$ and $u\to u+2c_v\,v$ with a number $c_v$. The sum $u+\Omega\,v$ is single-valued if and only if $c_v=\Omega$. We obtain $c_v$ from a numerical continuation of the system around the point, and it agrees with the value of the integral in eq.~\eqref{eq:boxclosed} to 36 digits. The continuation around a root of $16X^4-10X^2+3$ gives the same number.

\subsection{The expansion at large site energy}
\label{app:largeX}

We now derive the leading coefficients of eq.~\eqref{eq:boxlargeX}. At large $X$ the loop integral receives contributions from two regions, which we separate in $d=3-2\bar\eps$ dimensions. The regulator $\bar\eps=-\eps$ has the opposite sign to that of eq.~\eqref{eq:V}. In the soft region the loop momentum is of order one. We expand the integrand of the correlator in $y_e/X$,
\begin{equation}
\frac{1}{y_{12}\,y_{23}\,y_{34}\,y_{41}}\;\sum_{m\ge0}X^{-4-m}\,\kappa_m(y)\,,\qquad \kappa_0=1\,,\qquad \kappa_1=-\tfrac32\,\bigl(y_{12}+y_{23}+y_{34}+y_{41}\bigr)\,,
\label{eq:softexpansion}
\end{equation}
where $\kappa_m$ is a polynomial of degree $m$. In the hard region the loop momentum is of order $X$. The integrand is homogeneous of degree $-8$, so that this region starts at order $X^{-5-2\bar\eps}$. Poles in $\bar\eps$ cancel between the two regions and leave logarithms of $X$.

The leading term comes from the soft region alone, and it is $\Omega\,X^{-4}$ with $\Omega$ defined in eq.~\eqref{eq:boxclosed}. Its numerical value is $\Omega=21.4444295139357745195031667653426556$. To reduce $\Omega$ we introduce one Feynman parameter $x_e$ for each factor $1/y_e=1/|\vec\ell-\vec A_e|$, where the $\vec A_e$ are the vertices of the regular tetrahedron. The denominator becomes
\begin{equation}
\sum_e x_e\,(\vec\ell-\vec A_e)^2 \;=\; \Bigl(\vec\ell-\sum_ex_e\vec A_e\Bigr)^{2}+\sum_{e<f}x_e\,x_f
\label{eq:completesquare}
\end{equation}
on the simplex $\sum_ex_e=1$, because all distances between the vertices equal one. The Feynman-parameter formula for four factors of power one half has the prefactor $1/\pi^2$. With $\int\dd^3\ell\,(\ell^2+\Delta)^{-2}=\pi^2\,\Delta^{-1/2}$ the integral over $\vec\ell$ gives
\begin{equation}
\Omega \;=\; \int_{\sum x=1}\frac{\dd^3x}{\sqrt{x_1x_2x_3x_4}\;\sqrt{E_2(x)}}\,,\qquad E_2(x)=\sum_{e<f}x_e\,x_f\,.
\label{eq:Omegasimplex}
\end{equation}
The integral over the simplex can also be reduced to one dimension. With $\varrho=x(1-x)$ and $h=1+6\varrho$ we define
\begin{equation}
m_\pm\;=\;\frac12\Bigl(1\pm\frac{x^4-(1-x)^4}{h^2}-\frac{8\varrho\sqrt{(1+2\varrho)(4+9\varrho)}}{h^2}\Bigr)\,,
\end{equation}
and find
\begin{equation}
\Omega \;=\; 2\sqrt2\,\pi^2\int_0^1\frac{\dd x}{\sqrt{h}}\;{}_2F_1\bigl(\tfrac14,\tfrac34;1;m_+\bigr)\;{}_2F_1\bigl(\tfrac14,\tfrac34;1;m_-\bigr)\,.
\label{eq:Omega1d}
\end{equation}
Each hypergeometric function is a complete elliptic integral of the first kind. A simpler form is given in eq.~\eqref{eq:Omegahyp} below. To derive it, we write $x_1=x\,\theta$, $x_2=x\,(1-\theta)$, $x_3=(1-x)\,\theta'$ and $x_4=(1-x)\,(1-\theta')$. The integral over $\theta$ and $\theta'$ at fixed $x$ is an Appell function,
\begin{equation}
\int_0^1\!\!\int_0^1\frac{\dd\theta\,\dd\theta'}{\sqrt{\theta(1-\theta)\,\theta'(1-\theta')\,E_2}}\;=\;\frac{\pi^2}{\sqrt\varrho}\;F_2\Bigl(\tfrac12;\tfrac12,\tfrac12;1,1;-\frac{x}{4(1-x)},-\frac{1-x}{4x}\Bigr)\,.
\label{eq:F2step}
\end{equation}
Two identities for Appell functions \cite{Erdelyi:1953} then give eq.~\eqref{eq:Omega1d}. The first is the quadratic transformation
\begin{equation}
F_2(a;b,b';2b,2b';\xi,\zeta)\;=\;\Bigl(1-\tfrac{\xi+\zeta}{2}\Bigr)^{-a}F_4\Bigl(\tfrac a2,\tfrac{a+1}2;b+\tfrac12,b'+\tfrac12;\frac{\xi^2}{(2-\xi-\zeta)^2},\frac{\zeta^2}{(2-\xi-\zeta)^2}\Bigr)\,,
\label{eq:bailey}
\end{equation}
and the second is the product formula
\begin{equation}
F_4\bigl(a,b;c,a+b-c+1;m_+(1-m_-),m_-(1-m_+)\bigr)\;=\;{}_2F_1(a,b;c;m_+)\;{}_2F_1(a,b;a+b-c+1;m_-)\,.
\label{eq:F4product}
\end{equation}

At order $X^{-5}$ the soft region contributes $-6\,T_3$, where $T_3$ is the integral of eq.~\eqref{eq:boxclosed} without the factor $1/y_{41}$. The steps that led to eq.~\eqref{eq:Omegasimplex} now give
\begin{equation}
T_3 \;=\; \int\dd^d\ell\;\frac{1}{y_{12}\,y_{23}\,y_{34}} \;=\; \pi^{-\bar\eps}\,\Gamma(\bar\eps)\int_{\sum x=1}\frac{\dd^2x}{\sqrt{x_1x_2x_3}}\;E_2(x)^{-\bar\eps} \;=\; \frac{2\pi}{\bar\eps}-2\pi\bigl(\gamma_E+\ln\pi\bigr)-J_1+O(\bar\eps)\,,
\label{eq:T3}
\end{equation}
with
\begin{equation}
\int_{\sum x=1}\frac{\dd^2x}{\sqrt{x_1x_2x_3}}=2\pi\,,\qquad
J_1=\int_{\sum x=1}\frac{\dd^2x}{\sqrt{x_1x_2x_3}}\;\ln E_2(x)=4\pi\ln3-8\pi\,.
\label{eq:J1}
\end{equation}
The pole of the soft region is therefore $-12\pi/\bar\eps$. The hard region has the opposite pole, multiplied by $X^{-2\bar\eps}$. The sum is finite and contains the logarithm $-24\pi\,X^{-5}\ln X$, which is the coefficient $\beta_5$ of eq.~\eqref{eq:boxlargeX}. The finite parts combine to
\begin{equation}
\alpha_5 \;=\; \frac{\pi}{12}\,\bigl(22+291\ln3-108\ln2\bigr)\,.
\label{eq:alpha5}
\end{equation}

The higher orders follow in the same way. Each soft term is a one-loop integral with simple poles only, so that no higher power of the logarithm appears. A logarithm requires an ultraviolet pole of a soft term, which exists only at odd orders, so that $\beta_n=0$ for even $n$. We find
\begin{equation}
\alpha_6=-\frac{95}{3}\,\pi+8\,\widetilde\Omega\,,\qquad
\beta_7=-\frac{158}{3}\,\pi\,,\qquad
\alpha_7=\pi\Bigl(\frac{22519}{360}+\frac{423}{8}\ln3-16\ln2\Bigr)\,,
\label{eq:higherorders}
\end{equation}
where $\widetilde\Omega=-3.3142572208468856411918120617$ is the dimensionally regularized value of the integral in eq.~\eqref{eq:boxclosed} with $y_{12}^{-1}$ replaced by $y_{12}$. In terms of eq.~\eqref{eq:boxv} the coefficients are $\alpha_n=a_n\,\Omega+u_n$, where $a_n$ and $u_n$ are the coefficients of $X^{-n}$ in $v$ and in $u$, and the $\beta_n$ come from $u$ alone. The next two logarithms are $\beta_9=-\tfrac{627}{10}\pi$ and $\beta_{11}=\tfrac{10247}{840}\pi$, which agree with the numerical values to 12 and 8 digits. The series converges for $|X|>2$, beyond the outermost singular point of the system. Both $\Omega$ and $\widetilde\Omega$ can be written as finite combinations of hypergeometric functions at rational arguments. With the double series
\begin{equation}
S_j(a,b)\;=\;\sum_{m,n\ge0}\frac{(1)_{m+2n}\,(j+1)_{m+n}\,(2j+1)_m\,(\tfrac12)_n}{\bigl[(j+\tfrac32)_{m+n}\bigr]^2\,(1)_n}\,\frac{a^m\,b^n}{m!\,n!}\,,
\label{eq:Sseries}
\end{equation}
one finds
\begin{equation}
\Omega\;=\;\frac{2\pi^3}{\sqrt3}\Bigl[{}_2F_1\bigl(\tfrac18,\tfrac38;1;\tfrac19\bigr)\Bigr]^2-4\pi\,S_0\bigl(\tfrac14,\tfrac1{16}\bigr)\,,\qquad
\widetilde\Omega\;=\;\tfrac34\,\Omega-4E\,,
\label{eq:Omegahyp}
\end{equation}
where $E$ is the Feynman-parameter integral
\begin{equation}
E\;=\;\int_0^1\!\dd x\int_0^1\!\dd\theta\int_0^1\!\dd\eta\;\frac{Q^{1/2}}{\sqrt{\theta(1-\theta)\,\eta(1-\eta)}}\,,\qquad
Q\;=\;x(1-x)+x^2\,\theta(1-\theta)+(1-x)^2\,\eta(1-\eta)\,,
\label{eq:Eint}
\end{equation}
and $\Omega$ is the same integral with $Q^{-1/2}$; $E=4.8493948390746791327\ldots$ is itself a combination of seven Appell functions $F_2$ at $(\tfrac14,\tfrac14)$ and of $S_1(\tfrac14,\tfrac1{16})$. These formulas follow from a double Mellin--Barnes representation of the Feynman-parameter integral, and they agree with the integrals to 60 and 29 digits. We have found no relation between $\Omega$ or $\widetilde\Omega$ and known constants such as values of the zeta function.

In fact, every coefficient of eq.~\eqref{eq:boxlargeX} is a rational combination of $\pi$, $\pi\ln2$, $\pi\ln3$, $\Omega$ and $\widetilde\Omega$. The expansions of the soft and hard regions determine the complete boundary data of the system at $X=\infty$: all constants of its Frobenius solution are rational combinations of the five numbers $\pi$, $\pi\ln2$, $\pi\ln3$, $\Omega$ and $\widetilde\Omega$. Consequently $\alpha_n$ is a rational combination of $\pi$, $\pi\ln2$ and $\pi\ln3$ for odd $n$ and of $\pi$, $\Omega$ and $\widetilde\Omega$ for even $n$, while $\beta_n$ is a rational multiple of $\pi$ for odd $n$ and vanishes for even $n$. The next coefficients are
\begin{equation}
\begin{aligned}
\alpha_8&=-\frac{11783}{81}\,\pi-3\,\Omega+22\,\widetilde\Omega\,,\qquad&
\alpha_{10}&=-\frac{339187}{720}\,\pi-7\,\Omega+\frac{466}{15}\,\widetilde\Omega\,,\\
\alpha_9&=\pi\Bigl(\frac{34688681}{129600}-\frac{244}{15}\ln2+\frac{60317}{960}\ln3\Bigr)\,.
\end{aligned}
\label{eq:higherorders2}
\end{equation}
The structure of the system also fixes the class of functions. The matrix of the system of dimension 28 is block triangular with 19 diagonal blocks (25 for the system of $V_4$): one block of size 3 with rational solutions, one of size 5, which is the companion system of the operator $\mathcal L_5$ of eq.~\eqref{eq:L5}, three of size 2, which are the Picard--Fuchs systems of the three elliptic curves of appendix~\ref{app:curves}, and blocks of size 1 whose solutions are rational functions times the square roots of the letters. The longest chain of blocks has length 11 (12 for $V_4$), so that $u$, and with it $C_4$ and $V_4$, are iterated integrals of depth at most 10 (11) over rational and algebraic kernels and the periods of the three elliptic curves and of the K3 surface.

\subsection{The wavefunction coefficient on the line}
\label{app:V4}

For the coefficient $V_4$ the integrand has no factor $1/\prod_ey_e$, and the expansion at large $X$ starts at $X^{-5}$,
\begin{equation}
V_4\;=\;\sum_{n\ge5}X^{-n}\bigl(\alpha^V_n+\beta^V_n\ln X\bigr)\,.
\label{eq:V4largeX}
\end{equation} Through $X^{-11}$ the coefficients are rational combinations of $\pi$, $\pi\ln2$ and $\pi\ln3$; the constants $\Omega$ and $\widetilde\Omega$ first enter at $X^{-12}$. The first logarithm appears at $X^{-11}$ with $\beta^V_{11}=-\tfrac{11}{21}\pi$. In the space of dimension 8 these conditions leave two solutions without logarithms at large $X$, both of which begin at $X^{-12}$. Two exact conditions determine their coefficients. The first is single-valuedness at $t=-1+1/\sqrt{10}$, which is the condition that determines $\Omega$ in the correlator. The second is the absence of a pole at $t=0$, where the loop integral is finite. With these conditions $V_4$ contains no undetermined constant, and it agrees with the numerical values to 32 digits. The conditions determine in particular the coefficients $\alpha^V_{12}$ and $\alpha^V_{14}$ of the expansion. The soft region gives the same two coefficients as integrals in dimensional regularization,
\begin{equation}
\alpha^V_{12}=16\int\!\dd^d\ell\;y_{12}\,y_{23}\,y_{34}\,y_{41}-\frac{11596}{5103}\,\pi\,,\qquad
\alpha^V_{14}=256\int\!\dd^d\ell\;y_{12}^3\,y_{23}\,y_{34}\,y_{41}-\frac{2718675343}{220449600}\,\pi\,,
\label{eq:alphaV}
\end{equation}
and the two determinations agree to 36 digits. Integration by parts reduces the two integrals to $\Omega$, $\widetilde\Omega$ and $\pi$,
\begin{equation}
\alpha^V_{12}=\frac{\Omega}{56}+\frac{19\,\widetilde\Omega}{105}-\frac{58466}{25515}\,\pi\,,\qquad
\alpha^V_{14}=-\frac{4\,\Omega}{105}+\frac{2038\,\widetilde\Omega}{1575}-\frac{115599847}{8817984}\,\pi\,,
\label{eq:alphaVexact}
\end{equation}
and these expressions agree with the coefficients determined by the system to 32 digits. As for the correlator, the soft and hard regions determine all 34 boundary constants of the system exactly, and every $\alpha^V_n$ is a rational combination of $\pi$, $\pi\ln2$ and $\pi\ln3$ for odd $n$ and of $\pi$, $\Omega$ and $\widetilde\Omega$ for even $n$, with $\beta^V_n$ a rational multiple of $\pi$ for odd $n$ and zero for even $n$. The odd coefficients continue as
\begin{equation}
\begin{aligned}
\frac{\alpha^V_{11}}{\pi}&=\frac{760664581}{137168640}-\frac{1605383}{209952}\ln2+\frac{69}{64}\ln3\,,\\[2pt]
\frac{\alpha^V_{13}}{\pi}&=\frac{67959374957557}{9777380659200}-\frac{1160027389}{340122240}\ln2+\frac{2506681}{645120}\ln3\,,\\[2pt]
\frac{\alpha^V_{15}}{\pi}&=\frac{26196178644661273}{1048624075699200}-\frac{490944815353}{314272949760}\ln2+\frac{24289259}{2027520}\ln3\,,\\[2pt]
\frac{\beta^V_{13}}{\pi}&=-\frac{197}{56}\,,\qquad \frac{\beta^V_{15}}{\pi}=-\frac{1297433}{110880}\,,
\end{aligned}
\label{eq:alphaVodd}
\end{equation}
and the series agrees with the numerical values of $V_4$ to 32 digits.

%% file: appC_derivation.tex
\section{Direct derivation of the closed form with one leg per site}
\label{app:derivation}

In this appendix, we sketch a derivation of eq.~\eqref{eq:closedform} from the integral~\eqref{eq:V} that uses neither the differential equation of section~\ref{sec:L0} nor numerical values. We set $X_v=P_v$ throughout and write $\xi_i=\qG-2X_i$ and $u_i=\xi_i/\qG$, as in eq.~\eqref{eq:closedformW}. Two of the three integrations can be carried out in closed form, after which every remaining integral is the integral of a rational function times a logarithm in one variable. The symbol of these integrals and the cancellation of the terms of lower weight involve large expressions (about $6\times10^4$ terms), which we computed exactly by computer algebra. The scripts are included in the ancillary files.

\subsection{Site sums and the first integration}
\label{app:siteSums}

As coordinates we use the sums of the two edge energies at each site,
\begin{equation}
s_1=y_{12}+y_{31}\,,\qquad s_2=y_{12}+y_{23}\,,\qquad s_3=y_{23}+y_{31}\,,
\label{eq:sitesums}
\end{equation}
so that the partial energy of site $v$ is $X_v+s_v$, and $\dd^3y=\tfrac12\,\dd^3s$. The polynomial $\Bk$ of eq.~\eqref{eq:CM} is quadratic in each $s_v$. We split $\Vc=U_{12}+U_{23}+U_{31}$, where $U_{12}$ is the contribution to eq.~\eqref{eq:V} of the first term of the recursion~\eqref{eq:recursion3}, in which the edge $y_{12}$ is cut, and $U_{23}$ and $U_{31}$ follow by cyclic relabelling of the sites. With eq.~\eqref{eq:chain}, the integrand of $U_{12}$ is
\begin{equation}
\frac{\Bk^{-1/2}\,y_{12}\,y_{23}\,y_{31}}{(X_1+s_1)(X_2+s_2)(X_3+s_3)(\qG+2y_{12})}\left[\frac{1}{X_2+X_3+s_1}+\frac{1}{X_1+X_3+s_2}\right].
\label{eq:U12}
\end{equation}

We first integrate over $s_3$. Writing $\Bk=c_2s_3^2+c_1s_3+c_0$, at fixed $s_1$ and $s_2$ the region $\Bk>0$ is the interval between the two roots of $\Bk$, with $c_2<0$, and the integral is elementary. As a function of $s_3$, the integrand~\eqref{eq:U12} is a polynomial of degree one plus two simple poles, so the integral gives three terms. The polynomial part gives a rational function of $s_1$ and $s_2$ divided by $\sqrt{-c_2}$, where
\begin{equation}
-2c_2 \;=\; \kappa(P)+4P_1^2s_2^2+4P_2^2s_1^2-4\bigl(P_1^2+P_2^2-P_3^2\bigr)s_1s_2\,,
\label{eq:c2}
\end{equation}
with $\kappa$ the K\"all\'en function of eq.~\eqref{eq:Tpm}. The pole $X_3+s_3=0$ gives its residue divided by $\sqrt{-\Bk}$ at $s_3=-P_3$. There $-\Bk$ is half the square of a polynomial in $s_1$ and $s_2$ (because $X_3=P_3$), so this term is rational. Finally, the pole $\qG+2y_{12}=0$ gives the residue of proposition~\ref{prop:residues}, eq.~\eqref{eq:residueform}, which is a rational function of $z_{23}=y_{23}^2$ and $z_{31}=y_{31}^2$, while $\Bk$ is quadratic in them on this plane. The elliptic curve~\eqref{eq:quartic} therefore never appears: the remaining integrals over this plane give logarithms.

\subsection{Tangent lines and the second integration}
\label{app:tangent}

After the second integration, the square roots of $-c_2$ and of $-\Bk$ on the plane survive only through their values at the poles and at the ends of the range. The discriminant of $\Bk$ in $s_3$ is
\begin{equation}
c_1^2-4c_0c_2 \;=\; -4\,\qG\,\xi_1\xi_2\xi_3\,\bigl(s_1^2-P_1^2\bigr)\bigl(s_2^2-P_2^2\bigr)\bigl((s_1-s_2)^2-P_3^2\bigr)\,,
\label{eq:discs3}
\end{equation}
and the six lines on which it vanishes, $s_1=\pm P_1$, $s_2=\pm P_2$ and $s_1-s_2=\pm P_3$, are tangent to the conic $c_2=0$. On each of them $-2c_2$ is the square of a linear function, for instance $-2c_2=(P_1^2+2P_1s_2+P_2^2-P_3^2)^2$ on the line $s_1=-P_1$. Four of these lines bound the region of integration, $s_1\ge P_1$, $s_2\ge P_2$, $|s_1-s_2|\le P_3$, and with one leg per site the poles $X_1+s_1=0$ and $X_2+s_2=0$ lie on the other two. Every square root that is evaluated on these lines is therefore rational. The only poles that do not lie on a tangent line are the partial energies of pairs of sites in eq.~\eqref{eq:U12}. Each term of the bracket contains one of them, and we integrate each term first along the direction of its own pole: the term with $X_1+X_3+s_2$ over $s_1$, and the term with $X_2+X_3+s_1$ over $s_2$. The pole that is not on a tangent line is then a pole of the outer integral only, and every square root left by the inner integration is evaluated on a tangent line. The term from the pole at $s_3=-P_3$ is rational, and on the plane $\qG+2y_{12}=0$ we integrate first over $z_{23}$, where $-\Bk$ is again a perfect square at the pole and at the ends of the range.

The result is that each $U_e$ is a sum of eight one-dimensional integrals (four terms, each over two ranges),
\begin{equation}
\int_a^b\dd v\,\Bigl[\,\sum_gK_g(v)\,\ln|A_g(v)|+R_0(v)\Bigr]\,,
\label{eq:pieces}
\end{equation}
where $K_g$ and $R_0$ are rational functions and each $A_g$ is a ratio of products of linear factors in $v$.

\subsection{The symbol}
\label{app:symbolU}

The terms of weight two in eq.~\eqref{eq:pieces} come from the simple poles of the $K_g$, through
\begin{equation}
\int\frac{\ln|v-c|}{v-p}\,\dd v \;=\; \ln|p-c|\,\ln|v-p|-\mathrm{Re}\,\mathrm{Li}_2\Bigl(\frac{v-p}{c-p}\Bigr)\,,
\label{eq:antider}
\end{equation}
which gives the symbol of each integral in closed form. The only poles that are not rational functions of the momenta are a pair $\pm\rho$ in the plane term, at which the argument of the logarithm equals one, and their contributions combine into rational letters. Individual integrals contain letters that are absent from eq.~\eqref{eq:closedform}, such as $X_1+X_3$, $X_2+X_3$, and the values of quadratic polynomials at the ends of the ranges. These cancel within each $U_e$: the quadratic letters between the two ranges of each term, and the linear ones between the plane, pole, and polynomial terms. Computed symbolically in the $X_v$, the result is that $U_{12}$ has the symbol of a sum of five terms,
\begin{equation}
\frac{U_{12}}{\sqrt2\,\pi}\;\simeq\;a_1\,\mathrm{Li}_2(u_1)+a_2\,\mathrm{Li}_2(u_2)+a_3\,\mathrm{Li}_2(u_3)+b_1\,l_1l_3+b_2\,l_2l_3\,,\qquad l_i=\ln(1-u_i)\,,
\label{eq:SU12}
\end{equation}
where $\simeq$ denotes equality of symbols and
\begin{equation}
\begin{gathered}
a_1=\frac{2\qG\,\xi_2+\xi_1\xi_3}{16\,\qG\,\xi_1\xi_2^2}\,,\qquad
a_2=\frac{2\qG\,\xi_1+\xi_2\xi_3}{16\,\qG\,\xi_1^2\xi_2}\,,\qquad
a_3=\frac{X_3\,(\xi_1\xi_2+X_3\,\xi_3)}{4\,\qG\,\xi_1^2\xi_2^2}\,,\\
b_i=\frac{X_i^2X_3^2}{\qG\,\xi_1^2\xi_2^2\xi_3}\,.
\end{gathered}
\label{eq:aibi}
\end{equation}
In $U_{23}$ the third site is site~1, and in $U_{31}$ the second site is site~1. The coefficient of $\mathrm{Li}_2(u_1)$ in the symbol of $\Vc/\sqrt2\pi$ is therefore $a_1+a_3+a_2$, with the arguments of the last two relabelled accordingly, which gives
\begin{equation}
\frac{X_1}{8\,\xi_1\xi_2\xi_3}+\frac{X_1^2X_2X_3}{\qG\,\xi_1\xi_2^2\xi_3^2}\,,
\end{equation}
its coefficient in eq.~\eqref{eq:closedform}. Similarly, the coefficient of $l_1l_3$ receives $b_1$ from $U_{12}$ and $b_2$ from $U_{23}$, which sum to
\begin{equation}
\frac{X_1^2X_3^2\,(\xi_1+\xi_3)}{\qG\,\xi_1^2\xi_2^2\xi_3^2}\;=\;\frac{2X_1^2X_2X_3^2}{\qG\,\xi_1^2\xi_2^2\xi_3^2}\,,
\end{equation}
using $\xi_1+\xi_3=2X_2$. The remaining coefficients follow by relabelling, so the symbol of $\Vc$ equals that of eq.~\eqref{eq:closedform}.

\subsection{Terms of lower weight and the constants}
\label{app:constants}

The same computation gives the terms of lower weight. No $U_e$ has a term of weight zero, and the terms of weight one of $U_{12}$ are
\begin{equation}
\frac{U_{12}}{\sqrt2\,\pi}\bigg|_{\text{weight one}} \;=\; -\frac{X_1\,l_1+X_2\,l_2}{8\,\qG\,\xi_1\xi_2}+\frac{X_3^2\,l_3}{4\,\qG\,\xi_1\xi_2\xi_3}\,.
\label{eq:Ulow}
\end{equation}
These do not vanish, but they cancel in the sum over the three terms: the coefficient of $l_1$ in $\Vc$ is $-X_1(\xi_2+\xi_3-2X_1)/(8\,\qG\,\xi_1\xi_2\xi_3)=0$, since $\xi_2+\xi_3=2X_1$, and likewise for $l_2$ and $l_3$.

The difference between $\Vc$ and eq.~\eqref{eq:closedform} therefore has a vanishing symbol and no terms of lower weight, so it is a combination, with constant coefficients, of the rational prefactors that multiply the functions of weight two in eq.~\eqref{eq:closedform} and in the $U_e$. These prefactors span a space of rational functions of dimension 15, with denominators that divide $\qG\,\xi_1^2\xi_2^2\xi_3^2$, and no nonzero element of this space is finite at all three degenerate triangles $\xi_i=0$. Both $\Vc$ and eq.~\eqref{eq:closedform} are finite there. For the closed form, this is because the bracket vanishes to second order (section~\ref{sec:closedform}). For $\Vc$, it is because with one leg per site every denominator of the integrand is bounded away from zero on the region of integration ($X_v+s_v\ge2P_v$, while the partial energies of pairs of sites and the factors $\qG+2y_e$ are at least $\qG$). The difference therefore vanishes, which fixes in particular the term proportional to $\pi^2$ in eq.~\eqref{eq:closedformW}. As a check, evaluating the antiderivatives at the ends of each range gives $\Vc$ in closed form at any rational triangle, and at 31 triangles this agrees with eq.~\eqref{eq:closedform} to a relative precision of $10^{-45}$.

\subsection{Several legs per site}
\label{app:severalLegs}
\label{app:coefficients}

With several legs per site, $X_v>P_v$, the poles $X_v+s_v=0$ no longer lie on the tangent lines, and square roots survive the second integration. Their values at the points where these lines meet are the ten radicands of eq.~\eqref{eq:closedformgeneral}. $R_0$ and the $R_e$ are values of $-2\Bk$ by definition (section~\ref{sec:offline}), and the other six are values of eq.~\eqref{eq:c2}: $T^\pm_{12}$ is $-2c_2$ at $(s_1,s_2)=(-X_1,\pm X_2)$, and the other $T^\pm_{ij}$ follow by permuting the sites. We have not carried the derivation through for $X_v>P_v$.

The ten rational coefficients of eq.~\eqref{eq:closedformgeneral} can nevertheless be obtained from the integrand. By the construction of section~\ref{sec:offline}, each square root has exactly one integrable symbol, so it suffices to compute one nonzero part of the symbol of $\Vc$ for each of them. We take the part whose first entry is $X_1+P_1$. The discontinuity of $\Vc$ at $X_1=-P_1$ is given by the residue of the integrand on the pole $X_1+s_1=0$, integrated over the region that this plane cuts out. On the cut we integrate over $s_3$ as in appendix~\ref{app:siteSums}, which gives terms $\pm\pi/\sqrt{-\Bk}$ from the poles in $s_3$ and a term with $\pi/\sqrt{-c_2}$ from the polynomial part, and then over $s_2$, whose range again ends on tangent lines. (On the planes $\qG+2y_e=0$ we integrate over the square of the edge energy that varies, as in appendix~\ref{app:tangent}.) Each pole of the $s_2$ integrand, and its behavior at large $s_2$, gives an odd logarithm whose coefficient is a residue of the integrand. For the seven square roots that involve site~1, the result agrees with eq.~\eqref{eq:closedformgeneral}, up to a common sign (the orientation of the discontinuity) that the limit $X\to P$ fixes. The residues are exact, and at two kinematic points the two parts of the symbol agree to 37 significant digits. The residues are multiples of the prefactors with coefficients that do not depend on the kinematics. In the site sums, the integrand of eq.~\eqref{eq:V} has the residue $-\qG/16$ at the point where the partial energies of the three sites vanish ($R_0$), and $+\qG/16$ where those of the two sites at the ends of $e$ and $\qG+2y_e$ vanish ($R_e$). The terms with $T^\pm_{ij}$ come from the polynomial part in $s_3$, whose double residue is $-\rho^-_{ij}/8$ on the line where the partial energies of sites $i$ and $j$ vanish, and $-\rho^+_{ij}/8$ on the lines where the partial energy of site $i$ or $j$ and that of the pair $\{i,j\}$ vanish. The remaining three coefficients follow by relabelling the sites. This replaces the fit of section~\ref{sec:offline}, but it still assumes the form of the function space found there.